\documentclass[twocolumn,twocolappendix]{aastex701}
\usepackage{amsmath}

\usepackage{orcidlink}

\newcommand{\ruwe}{\texttt{ruwe}}
\newcommand{\drv}{\Delta \textrm{RV}_{\textrm{Gaia}}}
\newcommand{\rvgaia}{v_{\textrm{Gaia}}}
\newcommand{\rverrgaia}{\sigma_{v_{\textrm{Gaia}}}}

\begin{document}

\title{A search for black holes with metal-poor stellar companions: I. Survey sample selection and single epoch radial velocity follow-up}

\author[orcid=0000-0002-6406-1924]{Casey Y. Lam}
\affiliation{Observatories of the Carnegie Institution for Science, 813 Santa Barbara St., Pasadena, CA 91101, USA}
\email[show]{clam@carnegiescience.edu}  

\author[orcid=0000-0002-4733-4994]{Joshua D. Simon}
\affiliation{Observatories of the Carnegie Institution for Science, 813 Santa Barbara St., Pasadena, CA 91101, USA}
\email[hide]{jsimon@carnegiescience.edu}  

\author[orcid=0000-0002-6871-1752]{Kareem El-Badry}
\affiliation{Department of Astronomy, California Institute of Technology, 1200 E. California Blvd., Pasadena, CA 91125, USA}
\email[hide]{kelbadry@caltech.edu}  

\author[orcid=0000-0002-0531-1073]{Howard Isaacson}
\affiliation{Department of Astronomy, University of California, Berkeley, CA 94720, USA}
\email[hide]{hisaacson@berkeley.edu}  

\author[orcid=0000-0003-4727-4327]{Daniel D. Kelson}
\affiliation{Observatories of the Carnegie Institution for Science, 813 Santa Barbara St., Pasadena, CA 91101, USA}
\email[hide]{dkelson@carnegiescience.edu}  

\author[orcid=0000-0001-9611-0009]{Jessica Lu}
\affiliation{Department of Astronomy, University of California, Berkeley, CA 94720, USA}
\email[hide]{jlu.astro@berkeley.edu }  

\begin{abstract}
Stellar-mass black holes (BHs) above $30 M_\odot$ are predicted to form from low-metallicity progenitors, but direct detections of such systems in the Milky Way remain scarce.
Motivated by the recent discovery of Gaia BH3, a $33 M_\odot$ BH with a very metal-poor giant companion, we conduct a systematic search for additional systems.
Approximately 900 candidates are identified with Gaia as having significant deviations from single-star astrometric motion, evidence of RV variability, and low metallicities inferred from Gaia XP spectra.
We obtain single epoch high-resolution spectra for over 600 of these sources with Magellan/MIKE and Lick/APF and measure independent RVs with $\approx 1$ km s$^{-1}$ precision.
After removing contaminants such as hot stars, pulsators, eclipsing binaries, and hierarchical triples, we identify about 15 promising candidates with large RV amplitudes or offsets from the Gaia reported values.
This program establishes a well-characterized sample of BH candidates for detailed orbital modeling once Gaia DR4 epoch astrometry and RVs are released in late 2026; multi-epoch RV follow-up is ongoing.
Together, the Gaia and ground-based data will place new constraints on the demographics of BHs with metal-poor companions and test theoretical predictions linking low metallicity to the formation of the most massive stellar remnants.
% \keywords{}
\end{abstract}

\section{Introduction}
\label{sec:intro}

\subsection{Observational evidence connecting massive black holes with metal-poor progenitors}

Astrophysical stellar-mass black holes (BHs) are born from the collapse of massive stars.
The masses of these BHs range from $\approx 2 M_\sun$ to $\gtrsim 100 M_\odot$.
BHs spanning this full mass range have only been discovered in compact object mergers outside our local universe via gravitational waves \citep{Abbott_GW190521:2020, Abbott_GW190814:2020}. 
In contrast, nearly all BHs in our Milky Way with mass measurements span a much narrower range of $5-20 M_\odot$, whether those BHs are in X-ray binaries \citep{Ozel:2010, Farr:2011}, detached binaries \citep{El-Badry:2023a, Chakrabarti:2023, El-Badry:2023b}, or isolated systems \citep{Lam:2023, Sahu:2025}.
It is hypothesized that the $\gtrsim 30 M_\odot$ BHs discovered via gravitational waves form at low metallicity \citep{Belczynski:2010}.
Because wind driven mass loss decreases with decreasing metallicity, BHs that form from low metallicity stars will be more massive \citep{Heger:2003}.

Recently, \cite{Panuzzo:2024} reported the discovery of Gaia BH3, a $33 M_\odot$ BH with a $0.76 M_\odot$ very metal poor [Fe/H] $= -2.56$ giant star in an 11.6 year orbit.
Gaia BH3 is the most massive Galactic BH known to date, and is similar in mass to BHs discovered via gravitational waves.
In addition, the very metal poor giant companion appears to be associated with the very metal poor ED-2 stellar stream \citep{Balbinot:2024}.
Gaia BH3 thus provides direct evidence of the connection between low-metallicity stellar progenitors and massive stellar-mass BHs.

\subsection{Searching for additional massive BH binaries in the Milky Way}

The discovery of Gaia BH3 itself was also notable.
Unlike the first two BHs discovered with Gaia \citep[Gaia BH1 and BH2,][]{El-Badry:2023a, Chakrabarti:2023, El-Badry:2023b}, Gaia BH3 was not listed in any of the Gaia Data Release 3 (DR3) non-single-star (NSS) catalogs and did not have an orbital solution.
Instead, Gaia BH3 was discovered by \cite{Panuzzo:2024} during the process of preparing and validating the upcoming Gaia DR4.

\begin{figure*}
    \centering
    \includegraphics[width=1.0\linewidth]{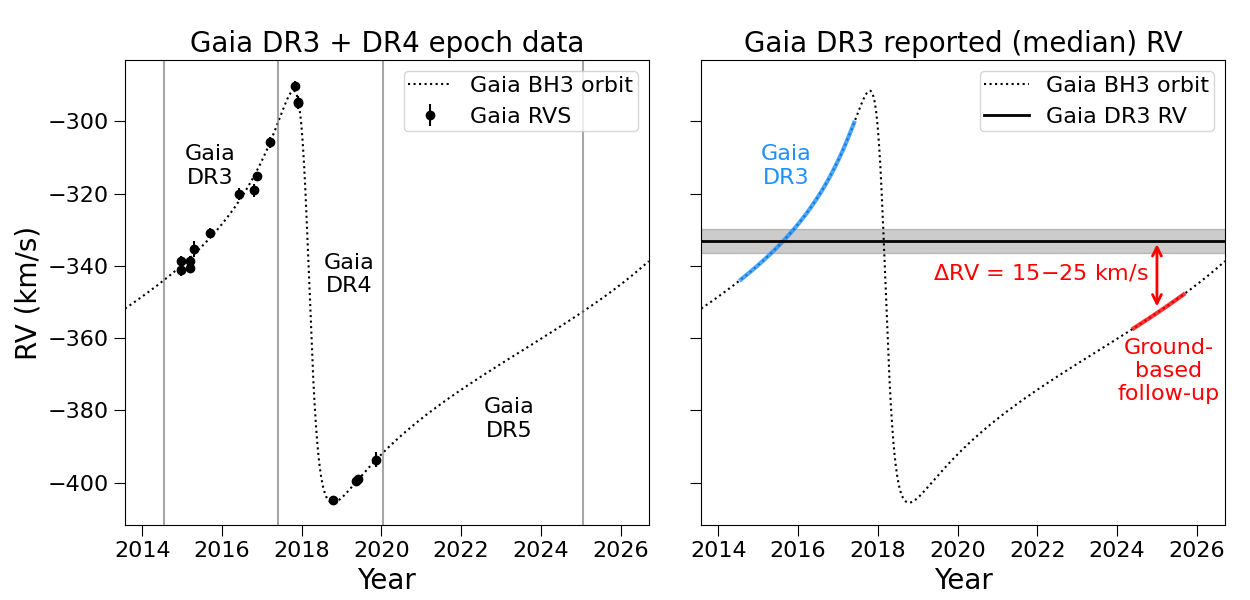}
    \caption{Left panel: RV measurements of Gaia BH3 taken with Gaia's Radial Velocity Spectrometer (black points) along with the orbital model determined by \cite{Panuzzo:2024} (dashed line).
    Gaia BH3's RVs in the Gaia DR3 window showed a roughly monotonic increase of $\approx 40$ km s$^{-1}$; RV quadrature occurred during the Gaia DR4 window, revealing the full RV amplitude of  $\approx 110$ km s$^{-1}$.
    Data in Gaia DR5 will nearly close the orbit.
    Right panel: In general, Gaia DR3 catalogs did not contain epoch RV measurements, only a median RV. 
    This is shown as the horizontal black line; the reported uncertainty is the shaded region.
    Ground-based follow-up observations of Gaia BH3 in 2024-2025 (red line) would be $15-25$ km s$^{-1}$ different from the Gaia DR3 median RV.
    This suggests large RV differences from the Gaia DR3 RV such that ground-based follow-up could be used to identify massive BHs in binaries with long orbital periods.
    \label{fig:gaia_bh3_rvs}}
\end{figure*}

Figure \ref{fig:gaia_bh3_rvs} illustrates why Gaia BH3 was discovered in Gaia DR4 but not Gaia DR3: because of its long orbital period, RV quadrature was only observable in DR4. 
Gaia DR3's temporal baseline is 34 months (2014--2017).
Gaia DR4, scheduled to be released 2026 December, has a temporal baseline of 66 months (2014--2020).

Even from the Gaia DR3 single-star catalog information, there were hints that Gaia BH3 was interesting: it showed astrometric motion inconsistent with a single-star motion model, as well as substantial radial velocity (RV) variation of $\approx 40$ km s$^{-1}$. 
Without an orbital solution, epoch astrometry, or RVs to validate the signal, it was not an obvious target of interest for follow-up (although see \cite{Muller-Horn:2025}, which presents a method for identifying compact objects using Gaia summary statistics in the absence of orbital solutions).
The discovery of Gaia BH3 suggests there are more massive BHs in the Galaxy waiting to be found, if we know where to look.
The sample of neutron stars and BHs discovered so far with Gaia tentatively suggests that low-metallicity stars have a higher occurrence rate of massive compact object companions than solar-metallicity stars \citep{El-Badry:2024a,El-Badry:2024b}.

In this work, we present ground-based RV measurements of Gaia BH3-like sources observed in 2024--2025.  
A single RV epoch taken in 2024 or 2025, compared to median RVs from 2014--2017 reported in Gaia DR3, will immediately enable detection of large RV variations potentially indicative of massive BH companions. 
Once the Gaia DR4 epoch data are released, the ground-based RV epoch can be simultaneously modeled with the Gaia data to provide independent validation of orbital solutions of BH candidates. 
If the orbital periods of BH candidates are long ($\gtrsim 10$ years), having observations now will reduce the amount of time required in the future to follow them up and validate their orbital solutions. 
A similar search with an independently constructed sample was conducted by \cite{Nagarajan:2025b}; a discussion and comparison are presented in Section \ref{sec:Discussion}.

The rest of this paper is organized as follows.
Section \ref{sec:Sample selection} defines the sample selection of our BH candidates.
Sections \ref{sec:Ground-based follow-up spectroscopy} and \ref{sec:Analysis of ground-based spectra} describe the acquisition and analysis of ground-based RVs, respectively.
Section \ref{sec:Results} presents the results of the follow-up, as well as an analysis of the systems that already have orbits mapped in Gaia DR3.
Section \ref{sec:Discussion} describes the contaminants in our sample, compares our sample to other searches for compact objects in the Gaia catalogs, and discusses the constraints this sample places on detached BH binaries.
We conclude in Section \ref{sec:Conclusions} and briefly describe the plan for continued follow-up of the most interesting candidates.

\section{Sample selection} \label{sec:Sample selection}

\begin{figure*}
    \centering
    \includegraphics[width=0.48\linewidth]{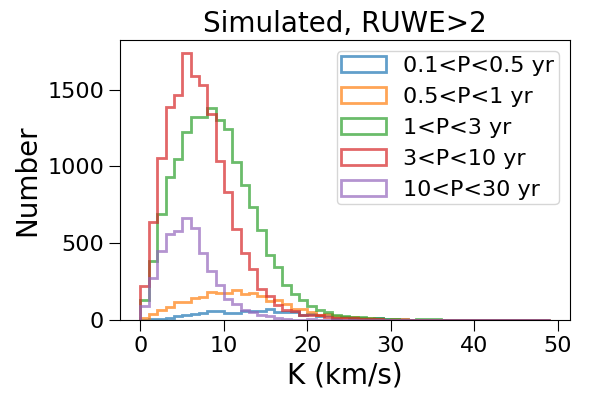}
    \includegraphics[width=0.48\linewidth]{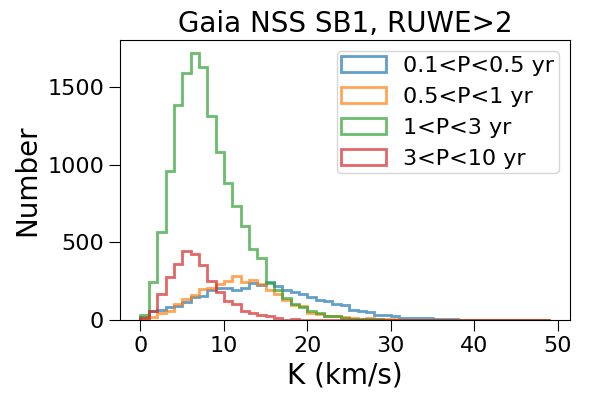}
    \caption{
    Right panel: RV semi-amplitude for star + star and star + WD binaries drawn from the population synthesis simulations of \cite{El-Badry:2024c}.
    Left panel: RV semi-amplitude of binaries drawn from the Gaia NSS SB1 catalog.
    For binaries where the orbital period is longer than 3 years, a threshold of K = 15 km s$^{-1}$ will eliminate most stellar binaries and improve the BH detection rate.
    The main contaminant will be binaries with shorter orbital periods; these can be ruled out with a few epochs of RV follow-up.
    The number of binaries within each period range differs between the left and right panels because the Gaia NSS processing pipeline is more sensitive to binaries with orbital periods $P < 3$ years, because the temporal baseline of Gaia DR3 is about 3 years.
    \label{fig:K}}
\end{figure*}

\begin{figure*}
    \centering
    \includegraphics[width=1.0\linewidth]{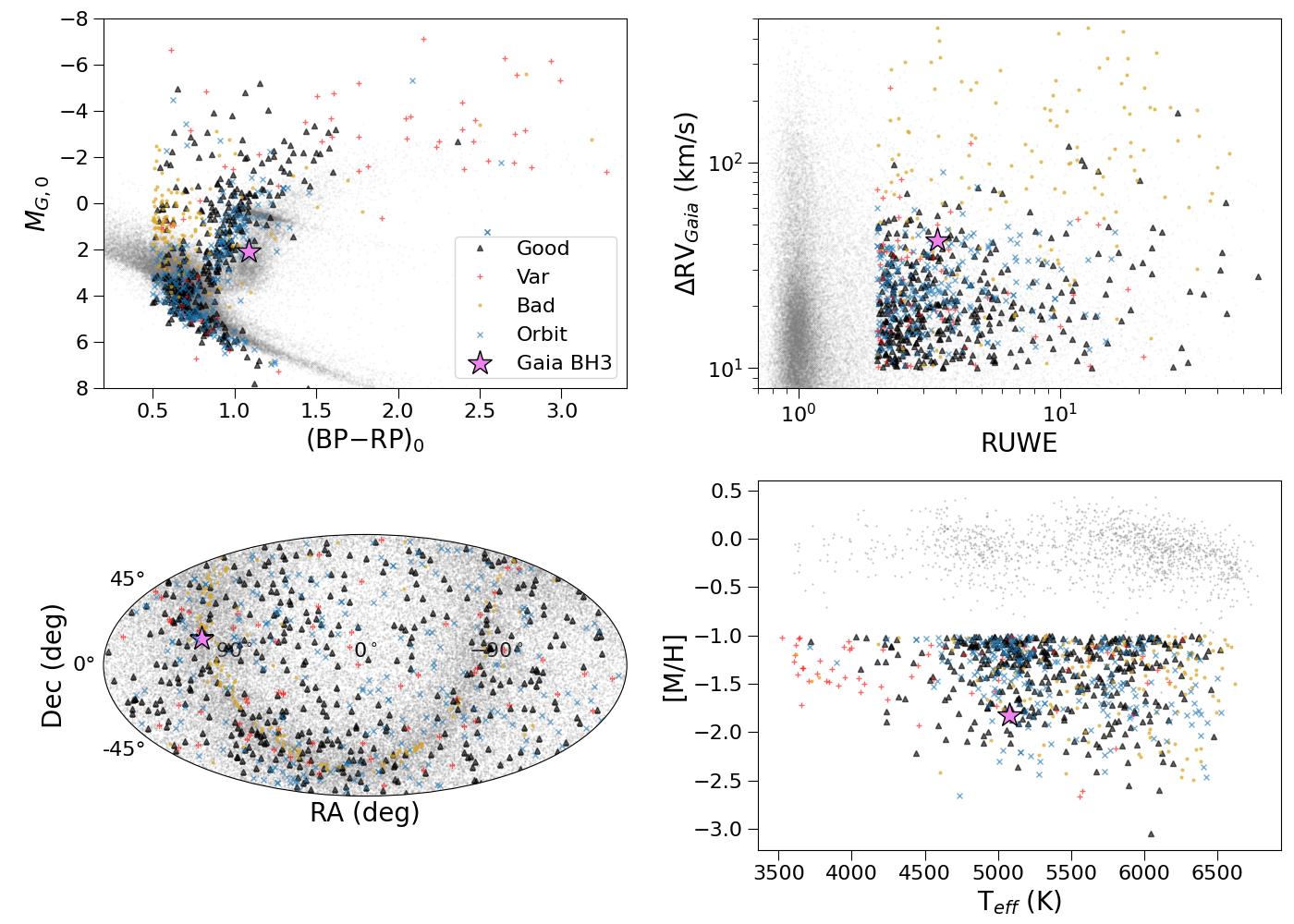}
    \caption{Targets, color-coded by their classifications from the observed RVs as discussed in \S \ref{sec:Results}.
    Sources with reliable RVs are ``Good" (black triangles), those that have reliable Gaia RVs but are classified as variable stars are ``Var" (red $+$), those that do not have reliably measured RVs are ``Bad" (yellow dots).
    Gaia BH3 is marked as the pink star.
    Sources that have orbital solutions from the Gaia NSS catalog are ``Orbit" (light blue $\times$) and we do not obtain RV follow-up of those sources.
    The gray points are a random subset of sources in the \texttt{gaia\_source} catalog that have $\varpi > 1$ mas, $\varpi/\sigma_\varpi > 5$, and \texttt{rv\_method\_used} = 1.
    Top left: extinction-corrected CMD. 
    Top right: $\drv$ vs. $\ruwe$.
    Bottom left: on-sky distribution.
    Bottom right: [M/H] vs. $T_{\textrm{eff}}$ inferred from the XP spectra, as reported in \cite{Andrae:2023b}.
    \label{fig:selected_classified}}
\end{figure*}

Inspired by the discovery of Gaia BH3, we search for massive BHs in wide binaries with low-metallicity companions.
In particular, we search for long-orbital period sources ($P_{orb} \gtrsim 1000$ days) which may have evaded inclusion in the Gaia DR3 astrometric and spectroscopic orbit catalogs.
We select sources that demonstrate significant deviation from single-star motion, have RV variation, are low metallicity, and are not hot. 
The selection cuts are as follows:

\begin{itemize}
\item Significant deviation from single-star motion ($\ruwe > 2$):
Gaia's renormalized unit weight error $\ruwe$ quantifies the goodness-of-fit of a single star astrometric model.
The $\ruwe$ is the square root of the reduced $\chi^2$ with an empirical correction to mitigate systematics in the $\chi^2$ distribution due to color and magnitude \citep{Lindegren:2018}.
Large $\ruwe$ (e.g., $ > 1.4$) indicates that a single star astrometric model is a poor fit; one possible explanation is that the source is a binary.
$\ruwe$ encodes information about the properties of the binary, such as the size of the photocenter's orbit, which is related to the orbital period of the binary \citep{Penoyre:2020, Lam:2025}. 

\item RV variation (\texttt{rv\_amplitude\_robust} $> 10$ km s$^{-1}$):
Gaia's \texttt{rv\_amplitude\_robust} ($\Delta \textrm{RV}_{\textrm{Gaia}}$) is the difference between the largest and smallest radial velocities $v_i$, after outlier removal.
Per \cite{Hambly:2022}, ``The statistical outliers are the valid transits in the time-series having RV values smaller than Q1 -- 3 $\times$ IQR or larger than Q3 + 3 $\times$ IQR, where Q1 and Q3 are the 25th and 75th percentiles, and where IQR = Q3 -- Q1."
Having detectable RV variation helps rule out false positives from large $\ruwe$ at very long orbital periods $P_{orb} > 10^{4.5}$ days (Figure \ref{fig:K}), due to the components being marginally resolved \citep{El-Badry:2024c}.
$\drv$ is only reported for sources where the magnitude of the flux integrated in the Radial Velocity Spectrometer (RVS) spectrum is $G_{\textrm{RVS}} < 12$.
The relationship between $G_{\textrm{RVS}}$ and $G$ depends on the color of the source, but the median difference between the two in our sample is $G - G_{\textrm{RVS}}\approx 0.8$ mag. 
This cut thus preferentially selects bright sources.

\item Low metallicity ($\textrm{[M/H]} < -1$):
\cite{Andrae:2023b} used machine learning techniques to infer metallicities for 175 million sources using low-resolution Gaia XP spectra.
The algorithm is trained on a sample of stars that span $-3 < \textrm{[M/H]} < 0.5$, and the [M/H] of \cite{Andrae:2023b} are more reliable than those released as part of Gaia DR3 in \cite{Andrae:2023a}.

\item Not hot ((BP--RP)$_0 > 0.5$):
Although improvements have been made to the Gaia DR3 pipeline for hot star RVs \citep{Blomme:2023},
RVs are still generally not reliable for hot and/or rapidly rotating stars.
We thus exclude them from our sample using a color cut.
Gaia BP--RP colors are extinction corrected using the 3-D dust map \texttt{mwdust} \citep{Bovy:2019}, which combines the dust maps of \cite{Marshall:2006}, \cite{Drimmel:2003}, and \cite{Green:2019}.
The BP--RP color excess is approximated as $\frac{4}{3}$E(B--V).
\end{itemize}
Applying these selection criteria results in a sample of 890 sources. 
Gaia BH3 is included in this sample ($\ruwe = 3.41$, $\drv = 42$ km s$^{-1}$, [M/H] = $-1.822$\footnote{[M/H] = $-1.822$ is from \cite{Andrae:2023a} as derived from the Gaia XP spectra, while the more accurate value of $\mbox{[Fe/H]} = -2.56$ reported in \cite{Panuzzo:2024} is derived using a high-resolution UVES spectrum obtained during follow-up.}, (BP--RP)$_0$ = 1.09\footnote{The \cite{Marshall:2006} dust map is likely under-predicting the extinction of Gaia BH3, making it look redder than it truly is; \cite{Panuzzo:2024} calculated (BP--RP)$_0$ = 0.92 for Gaia BH3.}).

% \startlongtable 
\tabletypesize{\small}
\begin{deluxetable}{ccccccc}
\tablecaption{Sample info.}
\label{tab:bh3like_sample}
\tablehead{\colhead{Gaia DR3 source ID} & 
    \colhead{$\ruwe$} & 
    \colhead{[M/H]} &
    \colhead{(BP$-$RP)$_0$} & 
    \colhead{RV (km s$^{-1}$)} &
    \colhead{$\Delta RV_{Gaia}$ (km s$^{-1}$)} & 
    \colhead{NSS solution}}
\startdata
\hline
16418496037121536 & 2.02 & -1.08 & 0.71 & 165.15 $\pm$ 4.56 & 49.41 & -- \\ 
31530905403904512 & 2.48 & -1.80 & 0.58 & -58.95 $\pm$ 1.94 & 20.89 & -- \\ 
34655678794913664 & 4.13 & -1.78 & 0.82 & -171.16 $\pm$ 1.34 & 11.28 & -- \\ 
67113609941350144 & 2.06 & -1.61 & 0.75 & -302.46 $\pm$ 1.39 & 21.11 & -- \\ 
71933624102004992 & 10.08 & -1.37 & 1.02 & -151.48 $\pm$ 2.59 & 30.44 & Orbital \\ 
109711129941111040 & 3.02 & -1.38 & 0.69 & -18.20 $\pm$ 1.57 & 18.56 & Acceleration7 \\ 
123015632954808576 & 2.60 & -1.70 & 0.90 & -147.15 $\pm$ 1.44 & 11.98 & -- \\ 
140001507056486656 & 2.37 & -1.70 & 1.07 & -212.10 $\pm$ 1.15 & 10.27 & -- \\ 
148926414737874048 & 2.84 & -1.12 & 1.18 & 91.13 $\pm$ 1.20 & 11.78 & SB1 \\ 
163873931722287360 & 3.23 & -2.03 & 0.65 & 220.09 $\pm$ 1.66 & 23.87 & AstroSpectroSB1 \\ 
\enddata
\tablecomments{Gaia DR3 source ID, \ruwe, RV, and $\drv$, and NSS solution come directly from the Gaia DR3 catalogs.
[M/H] is from \cite{Andrae:2023b}. 
(BP--RP)$_0$ is from the DR3 catalog, with extinction correction derived using \texttt{mwdust}.
This table is published in its entirety in machine-readable format.
A portion is shown here for guidance regarding its form and content.}
\end{deluxetable}

All candidates and their \ruwe, [M/H], (BP--RP)$_0$, Gaia-measured RV $\rvgaia$ and error $\rverrgaia$, $\drv$, and NSS solution (if applicable) are listed in Table \ref{tab:bh3like_sample}.
Note that Gaia defines $\rvgaia$ to be the median of all the epoch RVs, med($v_i)$, and that $\rverrgaia$ is defined as $\sqrt{\sigma_{med}^2 + 0.11^2} \, \textrm{km s}^{-1}$, where $\sigma_{med} = \sqrt{\frac{\pi}{2}} \frac{\sigma(v_i)}{\sqrt{N}}$ \citep{Hambly:2022}.

Figure \ref{fig:selected_classified} shows the CMD corrected for extinction using \texttt{mwdust}, on-sky distribution, $\drv$ vs. \ruwe, and [M/H] vs. $T_{\textrm{eff}}$ for all the candidates. 
The median magnitude of the sample is G=12.2, with 90\% of the sample having G=10.1 -- 13.1 mag.

This sample selection errs on the side of completeness over purity.
More stringent cuts on metallicity (e.g., [M/H]$ < -2$ and color (e.g., (BP--RP)$_0 > 0.7$) would have removed more unreliable sources and improved the number of good solutions. 
However, to avoid potentially missing an interesting target due to scatter or uncertainties in the parameter estimates, we accept a higher contamination rate. 

Out of the full sample of 890 sources, 586 sources, including Gaia BH3, do not have NSS solutions and are only found in the \texttt{gaia\_source} single-star catalog.
The other 304 sources are in at least one of the NSS catalogs, which includes solutions for astrometric orbits, spectroscopic orbits, joint astrometric-spectroscopic orbits, astrometric accelerations, and spectroscopic accelerations.
Some NSS sources have multiple types of NSS solutions because the processing for astrometric and spectroscopic solutions are independent. 
We describe the NSS properties in detail in Section \ref{sec:Results}.

\section{Ground-based follow-up spectroscopy}
\label{sec:Ground-based follow-up spectroscopy}

We obtain a single epoch of ground-based spectra to measure RVs for the sources without a measured orbital period in Gaia, i.e., sources that are only found in the \texttt{gaia\_source} catalog, or only have acceleration solutions in Gaia DR3.
This results in a target list of nearly 700 sources.
In the south, spectra were obtained with the Magellan Inamori Kyocera Echelle (MIKE) spectrograph  \citep{Bernstein:2003} on the 6.5m Magellan Clay telescope at Las Campanas Observatory.
In the north, spectra were obtained with the Levy spectrograph \citep{Vogt:2014} on the 2.4m Automated Planet Finder (APF) telescope at Lick Observatory.
We observed 617 sources between June 2024 and July 2025 using MIKE and APF; a summary of the observations is presented in Table \ref{tab:observations}.
We are actively following up the remainder of the sources and these will be presented in future work.

Since these sources are fairly bright, some have archival spectra from ground-based spectroscopic surveys.
We discuss archival spectra from several surveys (LAMOST, SDSS, RAVE) in Appendix \ref{app:Archival data}.

\tabletypesize{\small}
\begin{deluxetable*}{cccccccccc}
\tablecaption{Observation log.}
\label{tab:observations}
\tablehead{\colhead{Gaia DR3 source ID} & 
    \colhead{RA} & 
    \colhead{Dec} &
    \colhead{G} & 
    \colhead{Date} &
    \colhead{T$_{exp}$} &
    \colhead{SNR} & 
    \colhead{SNR} & 
    \colhead{SNR} & 
    \colhead{Inst.} \\
    \colhead{} & 
    \colhead{(deg)} & 
    \colhead{(deg)} &
    \colhead{(mag)} & 
    \colhead{} &
    \colhead{(sec)} &
    \colhead{(5175 \AA)} & 
    \colhead{(6563 \AA)} & 
    \colhead{(8525 \AA)} & 
    \colhead{}}
\startdata
\hline
1015566290114026752 & 134.933888 & 48.280852 & 12.33 & 2024-12-09 & 1200 & 24 & 29 & -- & APF \\ 
1030880524287375360 & 131.262116 & 55.452820 & 12.05 & 2024-10-02 & 1200 & 25 & 31 & -- & APF \\ 
1036998722380151296 & 133.784515 & 56.561330 & 12.34 & 2024-12-08 & 1200 & 19 & 21 & -- & APF \\ 
1052266579499163392 & 157.203362 & 62.994640 & 12.63 & 2024-09-28 & 1200 & 16 & 20 & -- & APF \\ 
1082902134545364608 & 117.033401 & 59.135673 & 12.56 & 2024-12-02 & 1800 & 6 & 6 & -- & APF \\ 
2332061310654863360 & 352.495839 & -27.419461 & 12.24 & 2024-07-28 & 300 & 40 & 94 & 105 & MIKE \\ 
2344504075725183104 & 13.110781 & -25.838040 & 12.37 & 2024-07-29 & 480 & 35 & 81 & 88 & MIKE \\ 
2347103974048275328 & 10.433787 & -24.885486 & 12.31 & 2024-07-29 & 600 & 41 & 92 & 97 & MIKE \\ 
2381708216313921920 & 350.112884 & -24.400393 & 11.91 & 2024-07-28 & 240 & 44 & 95 & 98 & MIKE \\ 
2392577197992227072 & 350.410294 & -20.310564 & 12.66 & 2024-07-28 & 360 & 36 & 82 & 87 & MIKE \\ 
\enddata
\tablecomments{SNR at 5175 \AA\ (around the Mg I b triplet), 6563 \AA\ (around H$\alpha$), and 8525 \AA\ (around the Ca II triplet) are listed.
This table is published in its entirety in machine-readable format.
A portion is shown here for guidance regarding its form and content.}
\end{deluxetable*}

\subsection{MIKE}

MIKE has a blue arm with 35 echelle orders spanning $3200 - 5000$ \AA, and a red arm with 34 echelle orders spanning $4900 - 10000$ \AA.
The two arms function as independent spectrographs.
Although observations were taken with both arms simultaneously, only data from the red arm are used to measure the RVs presented in this work.
Since we are systematics limited, not photon limited, including the blue arm would only introduce an additional source of systematic uncertainty when combining measurements from the two different spectrographs.
We choose to use the red arm because the SNR is higher and there are many absorption lines for making RV measurements \citep[including the Ca II triplet used by Gaia to measure RVs;][]{Cropper:2018}.
The MIKE observations were taken with the $0.7'' \times 5.0''$ slit,
resulting in a spectral resolution of $\sim 32,000$ in the red.

The typical exposure time at G=12.5 mag is 300 s, scaled up or down based on the source magnitude and observing conditions, resulting in moderate to high SNR spectra (median SNR per pixel $\approx 30$ at the Mg b I triplet, $\approx 80$ at H$\alpha$, and $\approx 90$ at Ca II triplet).

The observations were reduced with the MIKE pipeline from the Carnegie Python Distribution \citep[CarPy; ][]{Kelson:2000,Kelson:2003}.
CarPy performs flat-fielding, wavelength calibration, fringe correction, sky subtraction, and cosmic ray rejection.
The final result is a 1-D spectrum for each echelle order. 

\subsection{APF}
APF has 79 echelle orders that span a wavelength range of $3700 - 10200$ \AA, although wavelengths redward of $\approx 7300$ \AA\, are increasingly affected by fringing.
Observations were taken with the $2'' \times 3''$  T decker, resulting in a spectral resolution of $\sim 80,000$ at $5120$ \AA.
We do not use the iodine cell for our observations.

The typical total exposure time at G=12.5 mag is 1500~s, scaled up or down based on the source magnitude, resulting in moderate SNR spectra (median SNR per pixel $\approx 25$ at the Mg b I triplet, and $\approx 30$ at H$\alpha$).
For longer exposures $\gtrsim 900$ s, we typically take multiple exposures.

The observations were reduced using the California Planet Survey pipeline \citep[CPS; ][]{Howard:2010}.
The final result is a 1-D spectrum for each echelle order.
An identical wavelength solution is assumed for all targets, owing to the stability of the instrument.
We post-process the CPS reduced data using a custom pipeline to remove cosmic rays and stack multiple exposures together. 

\section{Analysis of ground-based spectra} \label{sec:Analysis of ground-based spectra}

We manually vet each MIKE and APF spectrum by eye in order to classify them as hot and/or rapidly rotating sources where the Gaia RV would have been unreliable; we do not derive RVs for these bad sources.
For the remaining sources, we use a custom pipeline to measure the RVs.

\subsection{Removal of unreliable Gaia RVs \label{sec:Removal of unreliable Gaia RVs}}

Gaia RVs are measured using the RVS instrument; the processing of the RVs in Gaia DR3 is described in \cite{Katz:2023}.
The RVS has a resolving power of 11,500 over a narrow wavelength range of 845--872 nm, designed to specifically cover the NIR Ca II triplet \citep[849.8, 854.2, 866.2 nm;][]{Cropper:2018}.

The objective of this project is to compare $\rvgaia$ to an independent ground-based RV to look for differences between the two.
However, this requires $\rvgaia$ to be reliably measured.
One necessary (but not sufficient) condition is that the source have detectable absorption lines in the Ca II triplet region.
If a source has no detectable absorption features in the RVS wavelength range, there is no way for a reliable RV to be measured when cross-correlating the source with a stellar template.

When vetting our sources, if there is no Ca II triplet in the MIKE spectrum, we mark the RV as bad, because Gaia's RVS instrument is only sensitive to that region.\footnote{The Ca II triplet region can also contain Paschen lines; however, this is only the case for several of our sources due to our color selection.}
The APF spectra do not cover the Ca II triplet, but they are sensitive to the Mg b I triplet in the optical (5167.3, 5172.7, 5183.6 \AA).
The Ca II triplet and the Mg b I triplet are both strong in FGKM stars, and thus the Mg b I triplet can be used as a proxy for the Ca II triplet.
Thus, if the APF spectrum does not have a detectable Mg b I line, we assume it also does not have a detectable Ca II triplet, and mark it as a bad RV.
We do not include these sources further in our analysis, because if the Gaia RVS velocities are unreliable, then $\drv$ (which is one of our main selection criteria) is also unreliable.
137 spectra fall into this category and are labeled ``Hot and/or RR (rapid rotator)" in Table \ref{tab:rvs}.

\subsection{Radial velocity pipeline}

When measuring the RVs from MIKE spectra, we use the orders spanning the wavelength range 5000 -- 8900 \AA, excluding the orders that contain Na D (due to potential interstellar contamination), as well as the wavelength ranges 6800 -- 7700\AA\, and 7900 -- 8500\AA\, (due to telluric absorption).
For our RV template, we use a MIKE spectrum of HD 126053, a slightly metal-poor solar-like star ([Fe/H] $\approx -0.4$, spectral type G1V; 
\citealt{Stefanik:1999}).

When measuring the RVs from APF spectra, we use the orders spanning the wavelength range 4400 -- 6700 \AA, where we again exclude the order containing Na D.
For our RV template, we use an APF spectrum of Tau Ceti (HD 10700).
Like HD 126053, Tau Ceti is a slightly metal-poor solar-like star ([Fe/H] $\approx -0.5$, spectral type G8V; 
\citealt{Stefanik:1999}). 

Each echelle order is treated as an independent measurement; we do not stitch the orders into a single spectrum.
For each order within the wavelength ranges stated above, we determine the RV that minimizes the $\chi^2$ between the science order and the template order.
The sigma-clipped mean and standard deviation of all the RVs is calculated.
For the remaining $N$ good RV measurements with mean $\bar{x}$ and standard deviation $\sigma$, we take the final RV to be $\bar{x}$ with a statistical uncertainty of $\sigma/\sqrt{N}$.

For both the MIKE and APF observations, we assume a systematic error floor of 1 km s$^{-1}$ for our RVs \citep[e.g.,][]{Chakrabarti:2023}, due to uncertainty in the wavelength solution, slit mis-centering, and template mismatches.
The majority of our statistical errors are $\ll 1$ km s$^{-1}$; the RV precision is systematics limited in most cases.
Since we are interested in relatively large RV changes and a precision of $\approx 1$ km s$^{-1}$ is sufficient, we defer improvements to the RV precision to future work \citep[e.g.,][]{Simon:2007, El-Badry:2024d}.

All the steps above are automated and the pipeline generally returns reliable RVs for cool stars; however, it fails about 5\% of the time when the majority of orders have very weak absorption lines.
When this happens, we manually select ``good" orders to be used to calculate RVs.
This typically happens when the number of good orders is $\lesssim 5$.

In other cases where the statistical error is $\gtrsim 1$ km s$^{-1}$, this is generally due to a poor template match.
For example, most sources that are flagged as variable stars by Gaia tend to have poor template matches and correspondingly large RV uncertainties.
Again, since the RV changes we are interested in are at least 10 km s$^{-1}$, even poor template matches are acceptable for our purposes and we defer improvement to future work.

\section{Results} \label{sec:Results}

\subsection{Radial velocity follow-up}
\label{sec:Radial velocity follow-up}

% \startlongtable 
\tabletypesize{\small}
\begin{deluxetable}{ccccccccc}
\tablecaption{Radial velocities and classifications.}
\label{tab:rvs}
\tablehead{\colhead{Gaia DR3 source ID} & 
    \colhead{MIKE/APF RV} & 
    \colhead{\# orders} & 
    \colhead{Gaia -- MIKE/APF} &
    \colhead{Gaia phot.} & 
    \colhead{Flags} \\
    \colhead{} &
    \colhead{(km s$^{-1}$)} &
    \colhead{used} &
    \colhead{RV (km s$^{-1}$)} &
    \colhead{var. flag} &
    \colhead{}}
\startdata
\hline
1015566290114026752 & -59.20 $\pm$ 1.00 & 27 & 4.26 $\pm$ 2.06 & 0 & -- \\ 
1030880524287375360 & 26.42 $\pm$ 1.00 & 26 & 5.74 $\pm$ 2.34 & 0 & -- \\ 
1036998722380151296 & -57.02 $\pm$ 1.00 & 25 & 3.07 $\pm$ 1.72 & 0 & -- \\ 
1052266579499163392 & -179.96 $\pm$ 1.00 & 18 & -2.25 $\pm$ 14.65 & 0 & -- \\ 
1082902134545364608 & -- & -- & -- & 0 & Hot and/or RR \\ 
2332061310654863360 & 92.48 $\pm$ 1.00 & 13 & -5.47 $\pm$ 2.34 & 0 & -- \\ 
2344504075725183104 & -75.32 $\pm$ 1.00 & 11 & 1.60 $\pm$ 2.48 & 1 & -- \\ 
2347103974048275328 & 118.09 $\pm$ 1.00 & 18 & 3.25 $\pm$ 2.00 & 0 & -- \\ 
2381708216313921920 & -15.55 $\pm$ 1.00 & 13 & -2.45 $\pm$ 2.56 & 0 & -- \\ 
2392577197992227072 & 2.21 $\pm$ 1.00 & 19 & 1.30 $\pm$ 2.38 & 0 & -- \\ 
\enddata
\tablecomments{If \texttt{phot\_variable\_flag} is VARIABLE (NOT AVAILABLE), it is listed as 1 (0) in the table.
If the source is a hot star or is rapidly rotating and thus there are not good absorption lines to measure the radial velocities, the Flag is ``Hot and/or RR" and no MIKE/APF velocity is derived so the MIKE/APF RV, \# orders used, and Gaia -- MIKE/APF RV entries are left blank.
Flags with SB2 (SB3) denotes a double-lined (triple-lined) spectroscopic binary; since it is unclear which star the RV is being measured for, these also have the entries blank.
This table is published in its entirety in machine-readable format.
A portion is shown here for guidance regarding its form and content.
}
\end{deluxetable}

\begin{figure*}
    \centering
    \includegraphics[width=1.0\linewidth]{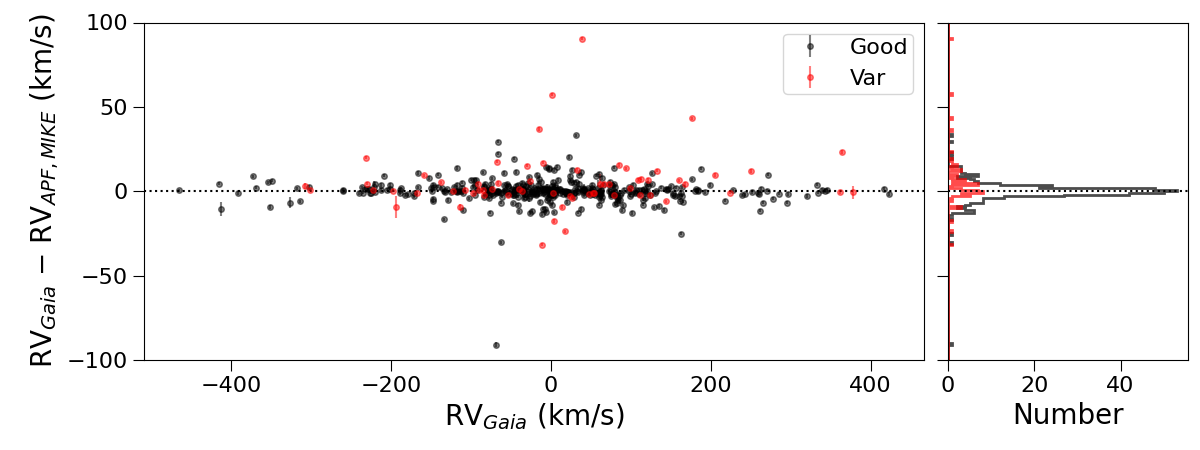}
    \includegraphics[width=1.0\linewidth]{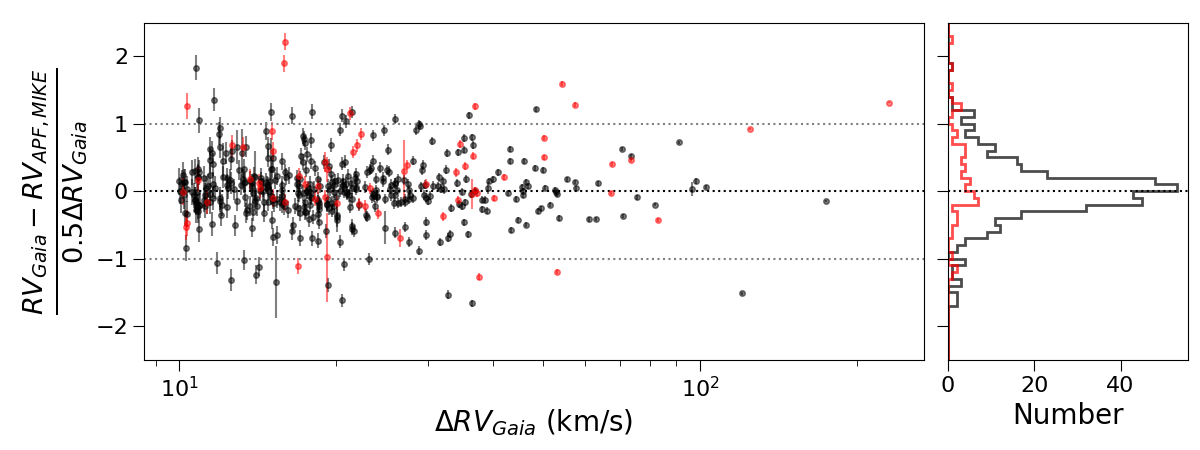}
    \caption{
    Results for the reliably measured RVs; sources that are variable stars are shown in red.
    Top row: Difference between $\rvgaia$ and the follow-up APF/MIKE RV (right), as a function of $\rvgaia$ (left panel).
    The y-error uncertainty is $\rverrgaia$ and the APF/MIKE RV error added in quadrature.
    Bottom row: Difference between $\rvgaia$ and the follow-up APF/MIKE RV, normalized by $0.5 \drv$ (i.e., the RV semi-amplitude robust; left), as a function of $\rvgaia$ (right).
    The y-error uncertainty is $\rverrgaia$ and the APF/MIKE RV error added in quadrature, normalized by $0.5 \drv$.
    Points that fall within the two gray horizontal lines are sources that have follow-up RVs that fall within Gaia's reported $\drv$, assuming that amplitude is centered on $\rvgaia$.
    In both the top and bottom panels, there are a few variable sources that fall outside the RV ranges of these plots.
    \label{fig:rvs}}
\end{figure*}

\begin{figure*}
    \centering
    \includegraphics[width=1.0\linewidth]{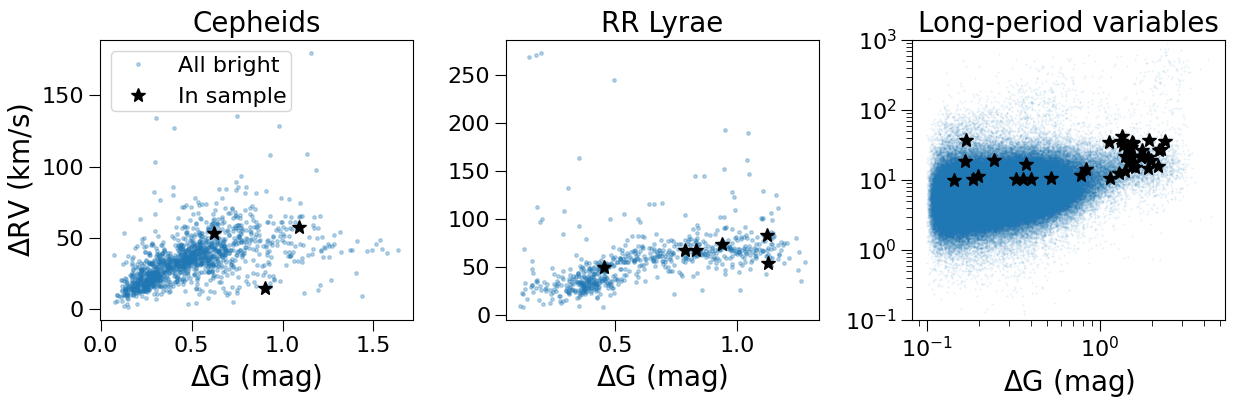}
    \caption{Gaia's \texttt{rv\_amplitude\_robust} ($\Delta RV_{Gaia}$) vs. photometric variability in Gaia DR3 $\Delta G$ for Gaia-identified Cepheids (left panel), RR Lyrae (middle panel), and long-period variables (LPVs, right panel).
    The Gaia variable source catalogs contain $1.5 \times 10^4$ Cepheids, $2.7 \times 10^5$ RR Lyrae, and $1.7 \times 10^6$ LPVs, however, Gaia RVs are only available for bright stars at $G \lesssim 13$, so the samples plotted are only 10\%, 0.3\%, 15\% of the Cepheids, RR Lyrae, and LPVs, respectively.
    The blue points show these measurements.
    The black stars show the overlap with our BH3-like sample. 
    The RV variation in our sample of Cepheids and RR Lyrae can all be explained by pulsations as they fall well within the roughly linear trend between RV and photometric variation; binarity is not needed to explain the RVs.
    LPVs typically have RV semi-amplitudes $\lesssim 10$ km s$^{-1}$ \citep{Arenou:2023}; $>99\%$ of the Gaia LPVs with RV measurements have $\Delta RV < 20$ km s$^{-1}$ which is consistent with this.
    LPV RV pulsational amplitudes do not show a strong dependence with $\Delta G$ nor $\ruwe$.
    LPVs are a heterogeneous class consisting of fundamental mode pulsators (Miras), long secondary periods, and ellipsoidal binaries \citep{Lebzelter:2023, Trabucchi:2023}.
    About half of our sample classified as LPVs have $\Delta RV > 20$ km s$^{-1}$, which potentially points to binarity but is difficult to definitively say.
    \label{fig:variables}}
\end{figure*}

Table \ref{tab:rvs} lists the results of the RV follow-up.
Of the sample of 616 stars, we were able to derive reliable RVs from our follow-up for 459 stars.
For the remaining 157 for which we could not derive an RV, the reason is noted in the table (i.e., hot and rapidly rotating).
Figure \ref{fig:rvs} shows the difference between $\rvgaia$ and the MIKE/APF RVs for the 459 reliable RVs.

Next, we consider the most interesting targets identified.
For systems without orbital solutions, we look for large RV amplitude and/or deviation from $\rvgaia$.
For systems with orbital solutions, we look for large mass functions.

\subsection{Interesting targets from radial velocity follow-up}
\label{sec:Interesting targets from radial velocity follow-up}

For the sources where their spectra suggest their RVs can be measured robustly, we now consider the sources that are the best BH candidates based on their RVs.

The RV semi-amplitude of the visible star $K$ is given by
\begin{multline}
    K = 30 \textrm{ km s}^{-1} \frac{M_2}{M_1 + M_2} \left( \frac{M_1 + M_2}{M_\odot} \right)^{1/3} \times \\ \left( \frac{P_{orb}}{\textrm{year}} \right)^{-1/3} \frac{\sin i}{\sqrt{1 - e^2}},
\end{multline}
where $M_1$ and $M_2$ are the masses of the primary and secondary, $i$ is the inclination, and $e$ is the eccentricity.
$K$ is an imperfect discriminator of massive binaries because it is a strong function of the inclination, which is a nuisance parameter.
In addition, without knowledge of the other parameters (especially $P_{orb}$), it is even more difficult to estimate what $K$ may correspond to interesting values of $M_2$.

\subsubsection{Large $\drv$}

Assuming that our sources have elevated $\ruwe$ because of astrometrically-detected unresolved binarity, it is unlikely that the orbital periods of these sources are very short, unless they are hierarchical triples.
Using the simulations of \cite{El-Badry:2024c}, we find that although sources with orbital periods $P_{orb} < 180$ d constitute $\sim 16\%$ of sources with $P_{orb} < 1000$ d, they constitute only $\sim 6\%$ of sources with $P_{orb} < 1000$~d that have $\ruwe > 2$.

Figure \ref{fig:K} suggests there will be few star + star and star + WD binaries with $K > 30$ km s$^{-1}$.
Keeping all other parameters fixed and assuming $M_1 = 1 M_\odot$, for $M_2 = 10 M_\odot$ vs. $M_2 = 1 M_\odot$, the value of $K$ will be more than 3 times larger.
Thus, $K > 30$ km s$^{-1}$ represents a reasonable selection criterion for BH companions, corresponding to $\drv \approx 60$ km s$^{-1}$, assuming a circular orbit with a period $1 \lesssim P_{orb}\lesssim 6$ years.

Table \ref{tab:rv_amp_60} lists the 20 RV-validated sources with $\drv > 60$ km s$^{-1}$.
6 (14) of the sources are (not) classified as photometrically varying in Gaia.

Although we have made robust RV measurements for all these sources, the four sources flagged with weak Ca II triplet are likely to have some unreliable Gaia measurements that inflate $\drv$.
In addition, these four sources have follow-up RVs that are consistent with $\rvgaia$ within the uncertainties.
This may suggest that although on average Gaia is able to measure their velocities, the scatter is very large and that the $\drv$ is entirely due to measurement scatter, and not physical velocity variations.
Additional follow-up can confirm or deny this hypothesis.

The four sources classified as RR Lyrae have RV variation consistent with pulsations (Figure \ref{fig:variables}) and do not show evidence of binarity.
The two eclipsing binaries also do not have a compact object companion.

This leaves 10 sources in Table \ref{tab:rv_amp_60} as viable candidates.
Of these, 7 sources have inconsistent $\rvgaia$ and MIKE/APF RVs;
% (Gaia DR3 5241421996088027392, 1824489875274815104, 1909404578470457216, 6196341900531496832, 3190523740098447744, 3329059422060781184, 2256759405698587136).
these are the best candidates for additional follow-up.
Several epochs of observations will be able to quickly rule out short-period contaminants and longer-term follow-up can characterize the most promising candidates.

\subsubsection{Moderate $\drv$ and MIKE/APF RVs different from $\rvgaia$}

Table \ref{tab:rv_amp_30_drv_15} lists the 18 RV-validated sources with $|RV_{APF/MIKE} - \rvgaia| > 15$ km s$^{-1}$ and $\drv > 30$ km s$^{-1}$.
Note that 7 of these sources are also in Table \ref{tab:rv_amp_60}.
These selection criteria are intended to identify Gaia BH3-like sources, i.e., sources that show RV deviations outside of the range they had during Gaia's time baseline. 

11 (7) of the sources are (not) classified as photometrically varying in Gaia.
Gaia DR3 1256697745659210368 has a 7-parameter astrometric acceleration solution and short-period SB1 solution ($P_{orb} = 30.7$ days). % APF
This suggests this system is likely a hierarchical triple.
The two eclipsing binaries rule out BH companions.
The four sources classified as RR Lyrae and the one Cepheid have RV variation consistent with pulsations (Figure \ref{fig:variables}) and do not show evidence of binarity.
V960 Mon underwent outburst in 2014 and was photometrically variable during the Gaia DR3 baseline; it has been identified as a FUor \citep{Semkov:2015}, and subsequently confirmed; it is also unlikely to have a BH companion.

This leaves 9 promising candidates, 3 of which were already identified in the previous section.
As discussed earlier, more spectroscopic follow-up is needed for these sources to determine their periods.
The exception is Gaia DR3 6143459892364148608, also known as RU Cen, which has a fully mapped orbit.

\begin{figure}
    \centering
    \includegraphics[width=1.0\linewidth]{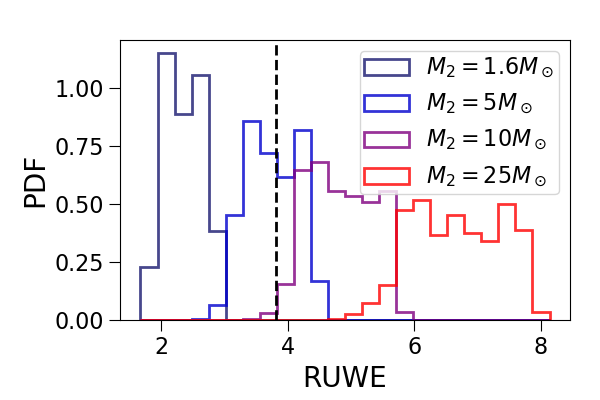}
    \includegraphics[width=1.0\linewidth]{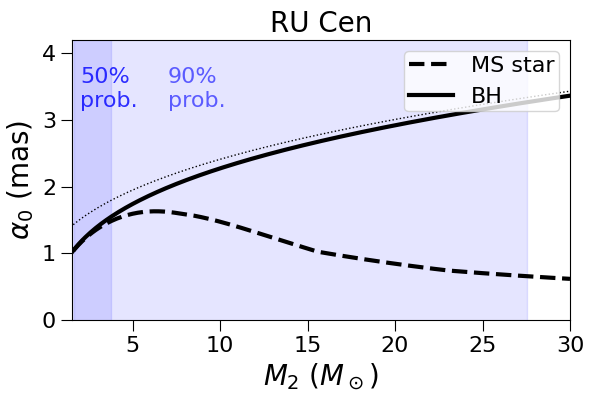}
    \caption{Top: Predicted $\ruwe$ for a variety of dark secondary masses $M_2$, predicted with \texttt{gaiamock}.
    The reported $\ruwe$ is shown in the dotted line.
    Note that this estimate of the $\ruwe$ is only based on the orbital properties, and does not account for variability, which may bias the reported $\ruwe$ upwards.
    Bottom: Predicted photocenter orbit semimajor axis size for RU Cen in mas, assuming a dark BH companion (solid line) or a luminous main sequence star (dashed line).
    For comparison, the thin dotted line shows the semimajor axis of the binary.
    Assuming an isotropic distribution of possible inclination angles $i$ (i.e., flat in $\cos i$), the darker (lighter) shaded region indicates the 50\% (90\%) probability range of the secondary mass $M_2$.
    The binary and orbital properties are $P_{orb} = 1489$ day, $e = 0.62$, $K = 22.1$ km s$^{-1}$, $M_1 = 0.6 M_\odot$, with the system at a distance of 2.3 kpc \citep{Oomen:2018}. 
    We use the mass-luminosity relation of \cite{Janssens:2022}, and assume that $M_G = -3.5$ for RU Cen.
    If the companion is $\lesssim 6 M_\odot$, it will be difficult to distinguish the nature of the companion using astrometry.
    \label{fig:ru_cen}}
\end{figure}

\subsubsection{RU Cen}

RU Cen is a post-AGB star, with $T_{\textrm{eff}} = 6000 \pm 250$K, $\log g = 1.5 \pm 0.5$, and [Fe/H] = $-2.0$ \citep{Maas:2002}.
It has an orbital period $P_{orb} = 1489 \pm 10$ days and eccentricity $e = 0.62 \pm 0.07$ \citep{Oomen:2018}.
RU Cen is not in the Gaia NSS catalog (i.e., no orbital solution nor classified as an acceleration sources).
Its orbital RV amplitude is $2K \approx 44$ km s$^{-1}$, but the amplitude of the pulsations at a given orbital phase is $\approx 5 - 15$ km s$^{-1}$; this makes deriving an orbital solution more difficult, because more epochs than usual are needed to disentangle RV variations from orbital motion vs. AGB pulsations \citep{Oomen:2018}.
The spectroscopic mass function (Equation \ref{eq:fMspec}) is $f_{M,S} =  (0.81 \pm 0.17) M_\odot$, suggesting it could be a good BH candidate.
However, it is difficult to determine the nature of possible low-mass companions to RU Cen, because of its high luminosity. 

We compare archival NUV data from GALEX \citep{Martin:2005} to the atmospheric model of RU Cen presented in Figure A.1. of \cite{Oomen:2018}. 
The NUV flux is $\lambda F_\lambda = 3.92 \textrm{ W/m}^2$, which is consistent with the atmospheric model.
Since we do not see a UV excess, this suggests that if there is a second luminous star in the binary, it is not very hot.

Although RU Cen does not have an astrometric solution in Gaia DR3, using the orbital parameters derived in \cite{Oomen:2018}, the expected $\ruwe$ as a function of the companion mass can be calculated.
The top panel of Figure \ref{fig:ru_cen} shows the predicted $\ruwe$ of RU Cen assuming a dark companion with different masses, as calculated using \texttt{gaiamock}, which simulates epoch Gaia astrometry and the Gaia processing pipeline \citep{El-Badry:2024c}.
The measured $\ruwe$ in Gaia DR3 is shown as the dashed vertical line; it is broadly consistent with a $5 M_\odot$ dark companion (although at this mass, a $5 M_\odot$ main sequence companion would be difficult to distinguish from a $5 M_\odot$ BH, see the next paragraph).
We also note that RU Cen is photometrically variable and that \texttt{gaiamock} does not simulate the effect of stellar variability on $\ruwe$.
Because the calculation of $\ruwe$ assumes that the source is not varying in color or magnitude, its value can be inflated for photometric variables \citep{Belokurov:2020}.
Thus, lower companion masses than suggested by the top panel of Figure \ref{fig:ru_cen} are also likely consistent with the observed $\ruwe$.
The distance is also uncertain ($2.3 \pm 0.5$ kpc, as derived from SED fitting and correcting for extinction assuming RU Cen has a luminosity of $L = 5000 L_\odot$ \citep{Gielen:2011}, as compared to 3.7 -- 7.7 kpc from simply inverting the Gaia DR3 single-star fit parallax), making it difficult to predict the expected photocenter orbit size.
In the calculation above, we have assumed the distance of 2.3 kpc.
Assuming distances of (1.8 kpc, 2.8 kpc) consistent with the $\pm 0.5$~kpc uncertainties leads to $\ruwe$ values which are (25\% larger, 15\% smaller).
Thus, in the case RU Cen is 2.8 kpc away, a $10 M_\odot$ BH companion would be most consistent with the observed $\ruwe$.

The bottom panel of Figure \ref{fig:ru_cen} shows the predicted size of the photocenter orbit semimajor axis, as a function of the unseen secondary's mass and whether it is a luminous main sequence (MS) star or dark compact object.
If the companion is $\lesssim 3 M_\odot$ (70\% probability, assuming an isotropic distribution of $\cos i$ and the $P, K, e$ reported in \citealt{Oomen:2018}), then the companion will contribute $\lesssim 1$\% of the optical light if it is a main-sequence star and $\delta_{ql}$ (see Section \ref{sec:Gaia DR3 NSS orbital solutions} and Equation \ref{eq:dql}) will also differ by $\lesssim 1$\% for the case of a luminous vs. dark companion.
This will make it effectively impossible to verify the nature of the companion in the optical.
If the companion is $\gtrsim 6 M_\odot$ (16\%), then the companion will contribute $\gtrsim 10$\% of the optical light if it is a main-sequence star, and $\delta_{ql}$ will also differ by $\gtrsim 10\%$, making it possible to distinguish a luminous vs. dark companion.
However, based on the top panel of Figure \ref{fig:ru_cen}, a companion mass $\gtrsim 10 M_\odot$ would be inconsistent with the observed $\ruwe$ if the source is closer than 2.8 kpc.
Spectral disentangling can be compared against astrometric measurements of the semimajor axis of the photocenter's orbit.
As discussed above, a major uncertainty is the distance to the system.
However, if the source is within 4 kpc and $M_2 > 6 M_\odot$, Gaia DR4 epoch astrometry should be able to determine the nature of the companion; a dark companion will have $\alpha_0 \gtrsim 2$ mas.

Another option is to try and directly resolve RU Cen's companion.
Assuming a distance of 2.3 kpc to the system, the semimajor axis of the orbit will be $>1.4$ mas (dotted line in Figure \ref{fig:ru_cen}).
The system is fairly bright (G=9) with a contrast $>10^{-3}$ if the companion is a main sequence star $\gtrsim 1.5 M_\odot$, making this a potential target for coronagraphy. 

RU Cen was observed by Hipparcos (HIP 59267).
However, RU Cen's Hipparcos goodness-of-fit statistic is 0.99, consistent with a single star model.
Hipparcos did not have the astrometric precision to detect RU Cen's binarity, and so we do not attempt to model the Hipparcos epoch data.

Finally, we note that RU Cen also resembles many previously identified ``BH impostors": a very luminous star with unusually low mass due to recent mass transfer \citep{Shenar:2020, Bodensteiner:2020, El-Badry:2022a, El-Badry:2022b, El-Badry:2022}.
Caution should be exercised in its analysis and interpretation.

\tabletypesize{\small}
\begin{deluxetable*}{ccccp{7cm}}
\tablecaption{RV-validated sources with $\Delta RV_{Gaia} > 60$ km s$^{-1}$}
\label{tab:rv_amp_60}
\tablehead{\colhead{Gaia DR3 source ID} & 
    \colhead{$\Delta RV_{Gaia}$} & 
    \colhead{$v_{r,Gaia}$} &
    \colhead{MIKE/APF RV} &
    \colhead{Comments} \\ 
    \colhead{} & 
    \colhead{(km s$^{-1}$)} & 
    \colhead{(km s$^{-1}$)} &
    \colhead{(km s$^{-1}$)} &
    \colhead{}}
\startdata
5352114607641222144 & 174.63 & -5.53 $\pm$ 34.74 & 6.87 $\pm$ 1.00 & Ca T very weak \\ % MIKE
5241421996088027392 & 120.16 & -68.44 $\pm$ 22.00 & 22.54 $\pm$ 1.54 & Ca T broadened but strong \\ % MIKE
4066043206990441600 & 102.44 & 2.09 $\pm$ 14.02 & -1.08 $\pm$ 1.50 & Ca T very weak \\ % MIKE
5786698704503419904 & 98.03 & -14.35 $\pm$ 9.22 & -21.76 $\pm$ 2.00 & NSS 7-parameter astrometric acceleration, Ca T broadened but strong \\ % MIKE
5618276028754440960 & 96.56 & 59.71 $\pm$ 6.12 & 57.95 $\pm$ 4.86 & Ca T very weak \\ % MIKE
1824489875274815104 & 90.84 & 31.93 $\pm$ 8.79 & -1.13 $\pm$ 1.00 & \\ % APF
6099072607957373056 & 81.93 & 1.18 $\pm$ 17.80 & 9.43 $\pm$ 1.00 & \\ % MIKE
5890573622432408960 & 75.57 & 2.39 $\pm$ 17.52 & 5.72 $\pm$ 1.42 & Ca T broadened and weak, emission in wings  \\ % MIKE
1909404578470457216 & 73.67 & -46.05 $\pm$ 3.45 & -65.40 $\pm$ 1.00 & NSS 9-parameter astrometric acceleration \\
6196341900531496832 & 70.96 & 102.12 $\pm$ 5.29 & 115.04 $\pm$ 1.00 & \\ % MIKE
3190523740098447744 & 70.60 & -65.92 $\pm$ 4.26 & -88.20 $\pm$ 1.00 & \\ % APF
437368717459759872 & 63.60 & -41.57 $\pm$ 9.63 & -45.24 $\pm$ 1.00 & \\ % APF
3329059422060781184 & 63.20 & -26.46 $\pm$ 6.80 & -13.62 $\pm$ 1.00 & \\ % APF, weird Ha
2256759405698587136 & 61.26 & -75.46 $\pm$ 2.56 & -62.70 $\pm$ 1.00 & \\ % APF
\hline
506869435873735552 & 229.88 & 34.02 $\pm$ 27.23 & -116.94 $\pm$ 1.05 & CS Per, eclipsing binary, $P_{orb} = 24.3$ days \citep{Samus:2017, Mowlavi:2023} \\ % http://www.sai.msu.su/gcvs/cgi-bin/search2.cgi?search=cs+per
4528862938544013184 & 124.47 & 0.95 $\pm$ 7.17 & -56.38 $\pm$ 1.00 & HD 342092, double-lined eclipsing binary with $P_{orb} = 4.11$ day \\
680218817055329920 & 83.06 & 4.37 $\pm$ 6.52 & 22.10 $\pm$ 1.00 & SS Cnc, RR Lyrae \\
1853751148171392256 & 73.68 & -10.28 $\pm$ 3.50 & -27.23 $\pm$ 1.00 & DM Cyg RR Lyrae, no evidence for binarity \citep{Barnes:2021} \\
3699778119760745472 & 67.76 & 93.89 $\pm$ 4.56 & 80.26 $\pm$ 1.00 & UV Vir, RR Lyrae \\
1798892252444818048 & 67.43 & 54.22 $\pm$ 5.92 & 55.20 $\pm$ 1.00 & BT Peg, RR Lyrae \\
\hline
\enddata
\tablecomments{Sources below (above) the horizontal line are (not) flagged as photometrically variable by Gaia.
Unless stated ``NSS", the sources are only in the \texttt{gaia\_source} catalog.
HD 342092 has not previously been identified as a variable star, but inspection of its TESS lightcurve \citep{Ricker:2015} shows it is a double-lined eclipsing binary with $P_{orb} = 4.11$ days.}
\end{deluxetable*}

\tabletypesize{\small}
\begin{deluxetable*}{ccccp{7cm}}
\tablecaption{$|RV_{APF/MIKE} - RV_{Gaia}| > 15$ km s$^{-1}$ and $\Delta RV_{Gaia} > 30$ km s$^{-1}$}
\label{tab:rv_amp_30_drv_15}
\tablehead{\colhead{Gaia DR3 source ID} & 
    \colhead{$\Delta RV_{Gaia}$} & 
    \colhead{$v_{r,Gaia}$} &
    \colhead{MIKE/APF RV} &
    \colhead{Comments} \\ 
    \colhead{} & 
    \colhead{(km s$^{-1}$)} & 
    \colhead{(km s$^{-1}$)} &
    \colhead{(km s$^{-1}$)} &
    \colhead{}}
\startdata
5241421996088027392 & 120.16 & -68.44 $\pm$ 22.00 & 22.54 $\pm$ 1.54 & Ca T broadened but strong \\ % MIKE 
2163305345451290240 & 36.41 & -62.59 $\pm$ 2.36 & -32.39 $\pm$ 1.00 & \\ % APF
1256697745659210368 & 48.44 & -66.33 $\pm$ 4.19 & -95.73 $\pm$ 1.00 & NSS 7-parameter astrometric acceleration solution and SB1 ($P_{orb} = 30.7$ days) \\ % APF
6657756868572776832 & 32.80 & 162.76 $\pm$ 3.37 & 187.86 $\pm$ 1.00 & \\ % MIKE
3190523740098447744 & 70.60 & -65.92 $\pm$ 4.26 & -88.20 $\pm$ 1.00 & \\ % APF
4203459792338061312 & 36.05 & 23.34 $\pm$ 6.88 & 3.09 $\pm$ 1.00 & \\ % APF
1909404578470457216 & 73.67 & -46.05 $\pm$ 3.45 & -65.40 $\pm$ 1.00 & NSS 9-parameter astrometric acceleration \\
\hline
506869435873735552 & 229.88 & 34.02 $\pm$ 27.23 & -116.94 $\pm$ 1.05 & CS Per, eclipsing binary, $P_{orb} = 24.3$ days \citep{Samus:2017, Mowlavi:2023} \\ % http://www.sai.msu.su/gcvs/cgi-bin/search2.cgi?search=cs+per
1823328795328595200 & 34.04 & 39.19 $\pm$ 3.25 & -51.19 $\pm$ 1.00 & RS Sge, RV Tauri b and eclipsing binary \citep{Samus:2017} \\ % NOTE: Not to be mixed up with RS Sgr, which has a lot of hits on ADS.
4528862938544013184 & 124.47 & 0.95 $\pm$ 7.17 & -56.38 $\pm$ 1.00 & HD 342092 \\
6262626680572363648 & 54.36 & 176.39 $\pm$ 4.58 & 133.16 $\pm$ 1.00 & VY Lib, RR Lyrae \\
6143459892364148608 & 57.55 & -15.08 $\pm$ 3.93 & -51.97 $\pm$ 1.00 & RU Cen, post-AGB, $P_{orb} = 1489$ day, $f_{M,S} = 0.81 M_\odot$ \citep{Oomen:2018} \\
203496585576324224 & 53.14 & -11.21 $\pm$ 6.40 & 20.44 $\pm$ 1.00 & SV Per, Cepheid, has nearby companion (0.2") which could result in spurious Gaia astrometry, no evidence for binarity \citep{Shetye:2024} \\
3101809648711790080 & 37.53 & 18.12 $\pm$ 4.48 & 41.90 $\pm$ 1.00 & V960 Mon, FUor \citep{Semkov:2015} \\
5288571906503118592 & 36.90 & 364.85 $\pm$ 2.25 & 341.66 $\pm$ 1.00 & ASAS J073325-6336.1, Mira variable \\
1925406252226143104 & 50.10 & -230.82 $\pm$ 4.95 & -250.75 $\pm$ 1.00 & AT And, RR Lyrae \\
680218817055329920 & 83.06 & 4.37 $\pm$ 6.52 & 22.10 $\pm$ 1.00 & SS Cnc, RR Lyrae\\
1853751148171392256 & 73.68 & -10.28 $\pm$ 3.50 & -27.23 $\pm$ 1.00 & DM Cyg RR Lyrae, no evidence for binarity \citep{Barnes:2021} \\
\hline
\enddata
\tablecomments{Sources below (above) the horizontal line are (not) flagged as photometrically variable by Gaia.
Unless stated ``NSS", the sources are only in the \texttt{gaia\_source} catalog.
Note that some entries are duplicated from Table \ref{tab:rv_amp_60} as the sources meet both criteria.
}
\end{deluxetable*}

\subsection{Gaia DR3 NSS orbital solutions
\label{sec:Gaia DR3 NSS orbital solutions}}

For sources that do not have an orbital solution, there is limited analysis that can be done because their orbital periods are unknown (although see \cite{Muller-Horn:2025} for an analysis using only Gaia summary statistics).
However, with an orbit mapped, much more information can be derived.
Different information about the binary is inferred depending on whether the orbit is mapped astrometrically or spectroscopically.

\begin{figure}
    \centering
    \includegraphics[width=1.0\linewidth]{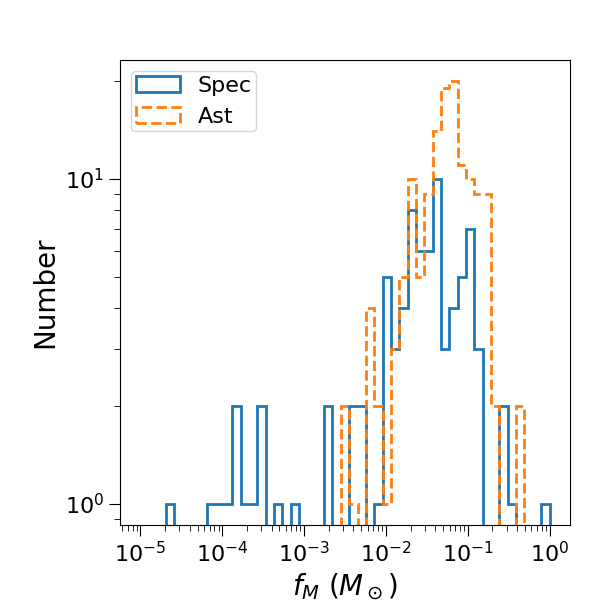}
    \caption{Mass function of the astrometric (Orbital), spectroscopic (SB1), and joint (AstroSpectroSB1) orbital solutions.
    Note that because some solutions were not merged, each solution is not necessarily a unique source.
    \label{fig:orbit_mass_function}}
\end{figure}

We first consider the astrometric case.
For a binary where the two components are unresolved, the photocenter's orbital semi-major axis in physical units $a_0$ is
\begin{equation}
    a_0 = a \delta_{ql}
\end{equation}
where $a$ is the semi-major axis of the binary, and 
\begin{equation}
    \delta_{ql} = \Bigg| \frac{q}{q+1} - \frac{l}{1+l} \Bigg|
    \label{eq:dql}
\end{equation}
where $q = M_2/M_1$ is the mass ratio and $l = L_2/L_1$ is the luminosity ratio; by definition, the primary (1) is the brighter, but not necessarily more massive, component in the binary. 
Thus, the photocenter orbital semi-major axis in angular units $\alpha_0$ is
\begin{equation}
    \frac{\alpha_0}{\textrm{mas}} = \frac{a_0}{\textrm{AU}} \frac{\varpi}{\textrm{mas}}.
\end{equation}

Because of the way $\delta_{ql}$ depends on both a binary's mass ratio and its luminosity ratio, luminous MS + MS star binaries and compact object + star binaries can be distinguished in $P_{orb} - a_0$ space.
At some fixed $P_{orb}$, a binary whose components are both luminous stars will have a smaller $a_0$ than a binary which has one component that is a compact object, because even though they have similar $a$, they will have very different $\delta_{ql}$ \citep{Shahaf:2019, El-Badry:2023a}.

The astrometric mass function is 
\begin{equation}
    f_{M,A} = \left( \frac{\alpha_0}{\varpi} \right)^3 \left( \frac{P_{orb}}{\textrm{year}} \right)^{-2} = (M_1 + M_2) \delta_{ql}^3 > M_1 + M_2,
\end{equation}
and enables a lower limit on the mass of the unseen secondary companion using astrometric observables, assuming the mass of the primary (visible) companion is known. 
It is a strong function of the photocenter orbit semi-major axis. 
It provides a more stringent constraint than the spectroscopic mass function (Equation \ref{eq:fMspec}).
This is because for MS + MS star binaries, $\delta_{ql} \lesssim 0.35$.
This means the maximum $f_{M,A} < 0.3 M_\odot$ for a sun-like star binary.
Thus, $f_{M,A} > 0.3 M_\odot$ indicates a compact object companion,  assuming the inferred orbital measurements are correct.

Next, we consider the case in which the orbit is only spectroscopically mapped, and no astrometry is available.
The spectroscopic mass function is
\begin{equation}
    f_{M,S} = \frac{P_{orb} K_1^3}{2 \pi G} \left(1 - e^2 \right)^{3/2} = M_2 \left( \frac{M_2}{M_2 + M_1} \right)^2 \sin^3 i < M_2,
    \label{eq:fMspec}
\end{equation}
and enables a lower limit on the mass of the unseen secondary companion using RV observables.
Because it is a strong function of inclination, a small mass function does not rule out a compact object companion.
However, a large mass function $f_{M,S} > 3 M_\odot$ with no evidence of a luminous secondary is best explained by a BH.

Our threshold for a potentially interesting solution is $f_{M,S}$ or $f_{M,A} > 0.3 M_\odot$, keeping in mind the probability of an interesting solution for this threshold is different for $f_{M,S}$ than it is for $f_{M,A}$.

For the targets in our sample which have Gaia SB1 solutions, we calculate the spectroscopic mass function $f_{M,S}$ using $P_{orb}$, $K_1$, and $e$ reported by Gaia.
For the targets in our sample which have Orbital and AstroSpectroSB1 solutions, we calculate the astrometric mass function $f_{M,A}$, where $P_{orb}$ and the Thiele-Innes components (which are used to calculate $a_0$) are reported by Gaia.
The histogram of the mass functions are shown in Figure \ref{fig:orbit_mass_function}.

There are four solutions with $f_M > 0.3 M_\odot$ (Table \ref{tab:orbital_solutions}); two solutions are uncombined SB1 and Orbital solutions, so there are actually only three unique sources with $f_M > 0.3 M_\odot$.
One of them (Gaia DR3 5136025521527939072, $f_{M,A} = 0.47 M_\odot$) is a previously identified NS candidate (see Section \ref{sec:Comparison to other compilations of compact object candidates}).
The remaining two candidates are Gaia DR3 197145394096715392 ($f_{M,S} = 0.88 M_\odot$) and Gaia DR3 2164592289448420224 ($f_{M,S} = 0.39 M_\odot, f_{M,A} = 0.48 M_\odot$) and have not been followed up in the literature. 

The SB1 model fit to Gaia DR3 197145394096715392 has a reasonable goodness-of-fit ($F_2$) and significance $s = K/\sigma_K$ (10.50). 
The $\Delta RV_{Gaia}$ and $K$ are consistent with each other given the predicted orbital period $P_{orb}$, suggesting the fit is plausibly correct.

The SB1 and Orbital model fit to Gaia DR3 2164592289448420224 are not consistent with each other, which would explain why they were not merged into an AstroSpectroSB1 solution.
The SB1 solution has a significantly better goodness-of-fit and higher significance ($F_2 = 1.3$ and $s = 38$) than the astrometric solution ($F_2 = 18$ and $s = 6.3$). 
For the SB1 solution, $\drv$ and $K$ are consistent with each other given the predicted orbital period $P_{orb}$, suggesting the fit is plausibly correct.
For the astrometric solution, the goodness-of-fit and significance suggest it to be unreliable.
It may be that during the orbit fitting process some type of minimum in the best $P_{orb}$ and $e$ was not correctly found, or that there was some problem with the astrometric measurements.
This suggests the astrometric mass function is slightly overestimated, but still passes our threshold for an interesting target.

\tabletypesize{\small}
\begin{deluxetable*}{c|cccc}
\tablecaption{Orbital solutions}
\label{tab:orbital_solutions}
\tablehead{\colhead{Source ID} &
    \colhead{5136025521527939072} & 
    \colhead{197145394096715392} & 
    \multicolumn{2}{c}{2164592289448420224}}
\startdata
Solution type & AstroSpectroSB1 & SB1 & Orbital & SB1 \\
$\drv$ (km s$^{-1}$)& 44.66 & 45.41 & 38.33 & 38.33 \\ 
$P_{orb}$ (day) & 536.90 $\pm$ 1.23 & 1023.04 $\pm$ 28.56 & 919.34 $\pm$ 98.78 & 711.08 $\pm$ 3.82 \\
$e$ & 0.65 $\pm$ 0.01 & 0.43 $\pm$ 0.09 & 0.05 $\pm$ 0.06 & 0.27 $\pm$ 0.02 \\
$a_0$ (mas) & 2.46 $\pm$ 0.05 & -- & 1.32 $\pm$ 0.21 & -- \\
$K$ (km s$^{-1}$) & -- & 22.71 $\pm$ 2.16 & -- & 18.26 $\pm$ 0.48 \\
$F_2$ & 4.68 & 2.24 & 18.10 & 1.31 \\
$s$ & 60.32 & 10.50 & 6.29 & 38.13  \\
\hline
\enddata
\tablecomments{Gaia DR3 5136025521527939072 is a NS candidate \citep{Arenou:2023,Shahaf:2023,El-Badry:2024b}.}
\end{deluxetable*}

\section{Discussion}
\label{sec:Discussion}

\subsection{Contaminants}

Extended/crowded sources, hot stars, variable stars, and hierarchical triples contaminated our sample or made it more difficult to identify good BH candidates.

\subsubsection{Extended/crowded sources}

Gaia DR3 3931816884728312320 is a galaxy (NGC 4569) and Gaia DR3 5248759101563602944 is in the core of a globular cluster (NGC 2808); we did not observe these two sources and they are not included in the analysis.
Gaia DR3 3931816884728312320 is classified as a galaxy in Gaia (\texttt{classprob\_dsc\_combmod\_galaxy} = 1.00), but made it into the sample with $\ruwe = 5.08$, (BP--RP)$_0$ = 0.93, $\drv = 303.91$ km s$^{-1}$, and [M/H] = --1.05.
These values are all spurious because the source is not a point source.
Gaia DR3 5248759101563602944 made it into the sample with $\ruwe = 2.49$, (BP--RP)$_0$ = 1.54, $\drv = 11.19$, and [M/H] = --1.14.
However, because it is in a crowded environment, its astrometry was contaminated; it has \texttt{ipd\_frac\_multi\_peak} = 14.00, which means 14\% of astrometric transits detected in Gaia has a double peak, suggesting a visually resolved double star.

\subsubsection{Hot stars}

Although hot stars' epoch astrometry and $\ruwe$ should be reliable indicators of binarity, $\drv$ is not; they are the primary contaminant of sources selected to have large $\drv$.
Despite a color cut, the sample contained a substantial number of hot stars.
In Figure \ref{fig:selected_classified}, most of the hot stars are in the Galactic Plane and have (BP--RP)$_0 < 0.6$ and $M_{G,0} < 4$.
Thus, although hot stars do not perfectly separate with a color-magnitude cut alone, adding a spatial criterion will help remove them more cleanly.
However, implementing such a cut would also have removed Gaia BH3, so in our preference for completeness over purity, we elected not to remove sources in the Plane.
Nevertheless, we emphasize that hot stars can only be reliably selected on the basis of their astrometry alone; spectroscopic follow up may be possible depending on the particular properties of that source.

\subsubsection{Photometrically variable stars}

About 10\% of the sample are variable stars, mostly long-period variables (LPVs). 
LPVs with large pulsations result in RV amplitudes that can easily detectable with our $\Delta RV_{Gaia} > 10$ km s$^{-1}$ threshold.
Even for variable stars that are in binaries, the RV signal will be composed of both binary and pulsational components.
Thus, the RV semi-amplitude of the binary due to orbital motion will be smaller than the total RV semi-amplitude.

In addition, many variable stars have large $\ruwe$ not because of binarity, but rather because the magnitude and color-dependent renormalization factor in calculating $\ruwe$ assumes the source is not photometrically variable.
If the assumption of a constant source magnitude is violated, $\ruwe$ will increase, even if the source does not have astrometric variability beyond parallax and proper motion \citep{Belokurov:2020}. 
Thus, although epoch astrometry should be reliable for isolated variable sources (since the centroid of the variable star does not change), the $\ruwe$ will be artificially inflated due to the photometric variability.
So although it is possible that there are variable star + BH binaries, the candidates in this sample are less likely to have BH companions than the non-variable sources, as their orbital RV amplitudes and astrometric signals are artificially inflated by their photometric variability.

Three of the LPVs were in the Gaia NSS catalog.
\cite{Arenou:2023} discuss how some SB1s can pick up the pulsational period incorrectly as an orbital period.
Two of the LPVs in the NSS catalog (612189661822368256 and 5299929895987315584) have NSS SB1 solutions where $P_{SB1} = 1/f_{LPV}$, suggesting the pulsations have been mistaken for orbital motion.
$P_{SB1} \approx 140$ days in both cases.
The remaining LPV in the NSS catalog does not have an orbital solution, and is only classified as a SecondDegreeTrendSB1. 
Since $1/f_{LPV} \approx 850$ days, it seems plausible that the pulsations have been mistaken for RV acceleration.

\subsubsection{Hierarchical triples}

Hierarchical triples will likely contaminate the sample (e.g., Gaia DR3 1256697745659210368).
A close inner spectroscopic binary will result in a large RV amplitude, while the wide outer astrometric binary will result in a large $\ruwe$.
Continued follow-up is needed to determine whether or not the sources have spectroscopic periods that match the periods inferred from astrometry.
Some hierarchical triple candidates are discussed in Appendix \ref{app:Additional details on non-single star solutions} and listed in Table \ref{tab:Potential hierarchical triples}.

\subsection{Comparison to other searches and compilations of compact object candidates}
\label{sec:Comparison to other compilations of compact object candidates}

Since Gaia DR3 was released in 2022 June, there have been many other searches for compact object candidates and catalog lists.
Here we list several compilations of candidates identified from the spectroscopic and astrometric orbits catalog.
Additional comparisons are listed in Appendix \ref{app:Additional cross-matches to compact object candidate catalogs and searches}.

\subsubsection{Comparison to sample of Nagarajan et al. 2025}

\begin{figure}
    \centering
    \includegraphics[width=1.0\linewidth]{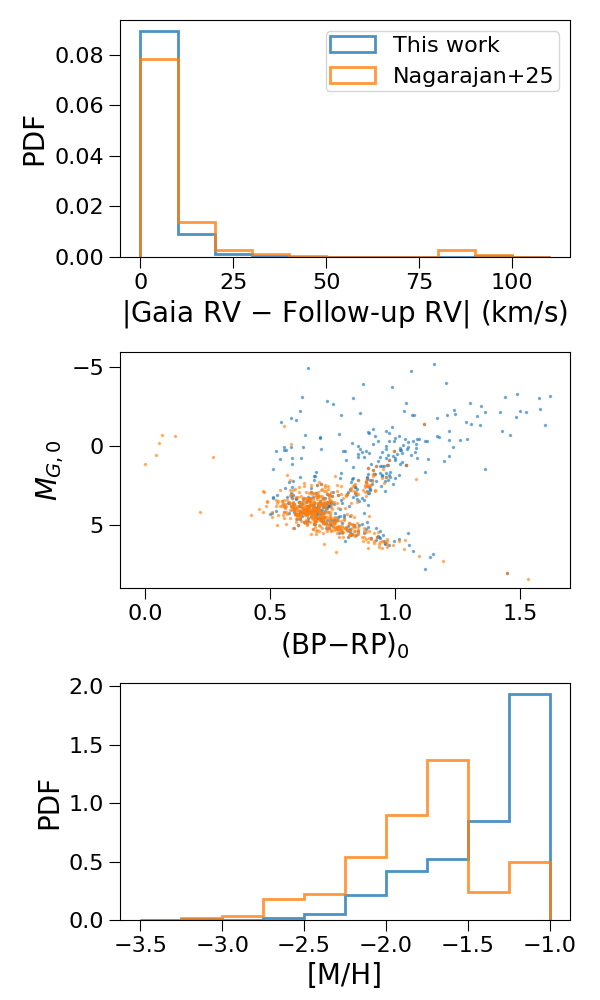}
    \caption{Comparison of the measured RV differences (top), CMDs (middle), and metallicity distribution (bottom) of the targets in this work and \cite{Nagarajan:2025b}.
    The measured RV differences across the two samples are similar.
    The sample presented in this work has substantially more giants than in \cite{Nagarajan:2025b}.
    The average metallicity in \cite{Nagarajan:2025b} is about 0.5 dex lower.
    \label{fig:comparison_to_pranav}}
\end{figure}

\cite{Nagarajan:2025b} also conduct a search for metal-poor stars with massive BH companions.
Their sample selection is qualitatively similar to ours (Gaia sources with RV measurements, metal-poor companions, cool stars, large $\ruwe$), but quantitatively different; see Section 3 of \cite{Nagarajan:2025b} for the details of their selection.
\cite{Nagarajan:2025b} has a sample of 887 sources; this work has a sample of 890 sources; 178 of the targets are in common, or about 20\% of the samples.

A major difference in the two samples' selection is that we include a criterion on $\drv$ for all sources, while \cite{Nagarajan:2025b} does not.
This means that on average, our sources are brighter, because only the brightest sources with $G_{RVS} < 12$ have a reported $\drv$ in Gaia DR3.
The median magnitude of our (the \citealt{Nagarajan:2025b}) sample is $G=12.2$ ($G=13.6$), and 90\% of the targets span $G = 10.1 - 13.1$ mag ($G = 12.0 - 14.5$ mag).

In Figure \ref{fig:comparison_to_pranav} we compare the measured RV differences, CMDs, and metallicity distributions of the two samples.
In these panels, we only consider the ``good" RVs from both samples; only sources in Table 3 of \cite{Nagarajan:2025b} that do not have a quality flag are included.

Although we had an explicit selection on RV variation and \cite{Nagarajan:2025b} did not, the distribution of the differences between the ground-based follow-up RVs and the Gaia RV is very similar across the two samples.
This suggests that selecting on $\drv$ may not be necessary in order to maximize observed RV differences.

Since the sample in this work is brighter than that of \cite{Nagarajan:2025b}, a larger fraction of the sources are giants (40\% vs. 10\%).
In addition, the sources in \cite{Nagarajan:2025b} are more metal-poor than the sources in this work, with a median [M/H] of $-1.73$ and $-1.27$, respectively.

\subsubsection{Previously identified candidates}

One of our sources with an orbital solution (Gaia DR3 5136025521527939072) is a neutron star candidate identified by \cite{Shahaf:2023} and Section 7.3 of \cite{Arenou:2023} as a NS/BH candidate and measured to have a secondary mass of $1.3 M_\odot$ in \cite{El-Badry:2024b}.
The candidate was identified using the method presented by \cite{Shahaf:2019}, which classifies astrometric binaries using the astrometric mass function divided by the primary mass, which they termed the ``astrometric mass ratio function ($\mathcal{A}$).

Six of our candidates with orbital solutions (Gaia DR3 272658547022867200, 2952986042608840064, 3379150938439271552, 4317453687509570688, 5136025521527939072, 5827340335106780928) are WD candidates in \cite{Shahaf:2025}. 
The \cite{Shahaf:2025} sample identified $\approx 3000$ WD binary candidates by looking for Gaia sources that had $\mathcal{A}$ that were poorly explained by MS + MS binaries, and did not show a red color excess indicative of a hierarchical triple stellar system.

Four of our candidates (Gaia DR3 1388975625910726272, 4897129549217063424, 6120659216659126016, and 6650410344195723264) are BH candidates in \cite{Muller-Horn:2025}.
Using Gaia DR3 summary statistics (e.g., $\ruwe$, \texttt{radial\_velocity\_error}), \cite{Muller-Horn:2025} identified 556 candidate red giant + BH binaries.
By construction, the catalog of \cite{Muller-Horn:2025} does not include any binaries that have an orbital period measured in the Gaia NSS.
\cite{Muller-Horn:2025} also construct a sample of 279 main sequence + BH binaries, although they caution this sample is likely significantly more contaminated than the red giant candidate list.
One of our candidates (Gaia DR3 5786698704503419904) is in this main sequence + BH candidate list.

\subsection{Population statistics}

Although only one RV epoch is not sufficient to confirm the nature of the BH candidates, we explore the population-level constraints that can be placed on detached BHs once the sources are classified.

First, we must quantify the selection effect of our sample construction.
We consider binaries with orbital periods $100 < P_{orb} < 10,000$ days.
We use the binary star + star and star + WD binaries drawn from the population synthesis simulations of \cite{El-Badry:2024c}.
For the compact objects, we simulate a population of $1 M_\odot$ stars + $10 M_\odot$ BHs and use \texttt{gaiamock} \citep{El-Badry:2024c} to predict their \ruwe.
For the BH binaries, we assume they spatially trace the stellar binary distribution, have a logarithmically flat period distribution, and a flat eccentricity distribution.
We calculate the expected $\drv$ and $\ruwe$ of all the sources.
To generate individual epoch RVs to calculate $\drv$, we assume that the Gaia RVs are sampled once every 40 days, over a time span of 1000 days, resulting in 25 RVs; this is a reasonable approximation of Gaia's scanning law, and roughly matches the median number of RV transits used in this sample.
We apply the outlier rejection used to calculate $\drv$ as described in the Gaia DR3 documentation.

Next, we impose our selection criteria of $\ruwe > 2$ and $\drv > 10$ km s$^{-1}$.
Our requirement for sources to have a reported $\drv$ in Gaia restricts the majority of sources to have $G < 13$, so we also impose this as a selection criterion.
We also explore the effect of making stricter selection on $\drv > 30$ and $\drv > 60$ km s$^{-1}$, since this is just taking a subset of our existing sample.

For the star + star and star + WD binaries, imposing the cuts $\ruwe > 2$ and $G < 13$ and $\drv > (10, 30, 60)$ km s$^{-1}$ will select (2, 0.3, 0.001\%) of sources within 2 kpc.
For the $1M_\odot$ MS star + $10 M_\odot$ BH binaries, imposing the cut $\ruwe > 2$ and $G < 13$ and $\drv > (10, 30, 60)$ km s$^{-1}$ will select (6, 5, 4\%) of sources within 2 kpc.
For reference, we also calculate our statistics for binaries containing giants, since our sample contains a significant fraction of them.
When imposing these cuts on giant star + star and giant star + WD binaries, they will select (12, 2, 0.01\%) of sources within 2 kpc.
For $1M_\odot$ giant star + $10 M_\odot$ BH binaries, imposing the cut $\ruwe > 2$ and $G < 13$ and $\drv > (10, 30, 60)$ km s$^{-1}$ will select (56, 48, 35\%) of sources within 2 kpc.

If the number of BHs in detached binaries per star is assumed to be $10^{-6}$ at solar metallicity across $100 < P_{orb} < 10,000$ day, and BHs at low metallicity are over-represented by a factor of 100 \citep{El-Badry:2024}, then the occurrence rate of BHs per metal-poor star is $\eta_{BH} = 10^{-4}$.
This means the number of BH+star:star+star binaries  will be (1:3300, 1:1500, 1:3) for the cut on $\drv > (10, 30, 60)$ km s$^{-1}$ for sources within 2 kpc.
The contamination rate is a very steep function of $\drv$.
This suggests that selecting for very large $\drv$ sources will significantly decrease the contamination rates.
The main difficulty in using $\drv$ as a selection criterion is the significant number of false positives, due to hot stars, pulsating variable stars, or hierarchical triples.
Hot stars and variable stars can be excluded by stricter color and variability cuts; hierarchical triples will be more difficult to rule out without some RV follow-up.
We will continue to follow-up the sources with $\drv > 30$ km s$^{-1}$, but we do not expect to have many detections until $\drv \gtrsim 50$ km s$^{-1}$.

Section \ref{sec:Interesting targets from radial velocity follow-up} presents a dozen promising sources with $\drv > 60$ km s$^{-1}$.
If there are $10^{-4}$ BHs per metal-poor star, then 4 should be BHs.
If the occurrence rate is an order of magnitude lower, then we would expect a 30\% chance of one of those sources being a BH. 
If no BHs are detected, occurrence rates $\eta_{BH} \gtrsim 10^{-6}$ can be ruled out.
We note that this prediction is strongly dependent on the accuracy of the reported $\drv$, as well as the assumption these binaries have orbital periods $100 < P_{orb} < 10,000$ days.
Additional RV follow-up is needed to validate these statements.

\begin{figure}
    \centering
    \includegraphics[width=1.0\linewidth]{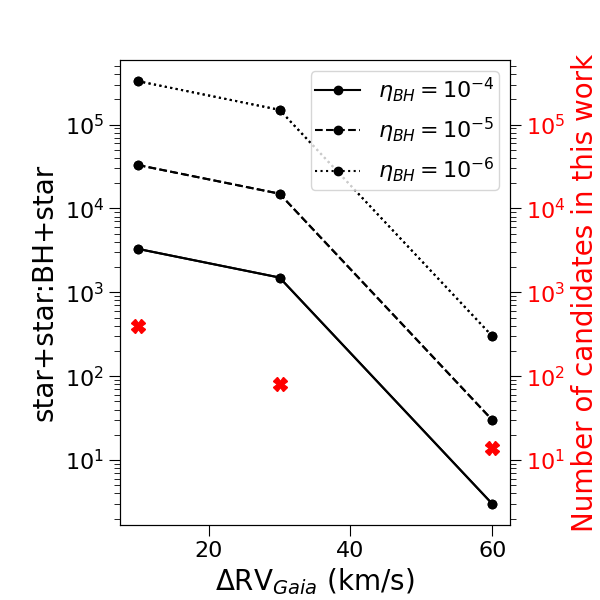}
    \caption{The ratio of star+star:BH+star systems with $\ruwe > 2$, $G < 13$, and $\drv > (10, 30, 60)$ km s$^{-1}$, as a function of the BH occurrence rate $\eta_{BH}$ (black curves).
    This can be interpreted as the number of star+star binaries needed to expect to have 1 BH detection.
    In order to estimate the potential BH yield of our sample, we compare this to the number of sources in our sample as a function of $\drv > (10, 30, 60)$ km s$^{-1}$ (red crosses).
    The ratio star+star:star+BH binaries as a function of $\drv$ is significantly steeper than the number of systems in our sample as a function of $\drv$.
    This suggests that the largest $\drv$ systems are the most promising.
    If the BH occurrence rate for metal-poor stars is $\eta_{BH} = 10^{-4}$, then we would expect to be able to detect a BH in our sample.
    \label{fig:bh_yield}}
\end{figure}

Understanding the detached BH binary population as a function of both orbital period and metallicity will require a significantly larger sample.
Previous work has placed constraints on the occurrence rate of BHs in wide binaries based on the detections of Gaia BH1, 2, and/or 3 \citep[e.g.,][]{El-Badry:2023a, Nagarajan:2025a, Lam:2025}.
However, even when selection effects are understood, it is difficult to place tight constraints on a rare population with only a few detections; the best existing constraints are limited to the orbital periods of Gaia BH1 and BH2 (roughly 0.5 years and 3 years, respectively).
Gaia DR4 will provide the sample size needed, with epoch RV and astrometry measurements enabling the detection of promising candidates.

\section{Conclusions}
\label{sec:Conclusions}

We present a survey of metal-poor stars with astrometric deviations from single-star motion in order to search for detached BHs with long orbital periods in Gaia DR3.
We obtain 1 epoch of high-resolution, moderate-to-high SNR spectra for over 600 sources in order to search for large RV variations potentially indicative of a massive BH companion.
About 12\% of the sample was contaminated with variable stars whose inferred $\ruwe$ are not reliable, and another 24\% of the sample was contaminated by hot and/or rapidly rotating stars whose RV measurements were not reliable.
Most of the remaining 64\% of the sources were found to be consistent with the Gaia RV expectations.
We are actively obtaining spectra for the final $\approx 10\%$ of our sample that has not been followed up, as well as multiple epochs of follow-up spectra to characterize the best candidates we identified in Section \ref{sec:Results}.
These results will be presented in a follow-up paper.
If the occurrence rate of BHs around metal-poor stars is between $10^{-5} - 10^{-4}$, then we expect to detect between 0.3 and 3 BHs in our current sample.
Future work will also include modeling the epoch astrometry and RV measurements from the upcoming Gaia DR4.

\medskip

The APF and MIKE spectra described in the publication of this paper will be archived in a Zenodo repository upon acceptance (link to be added).

\begin{acknowledgments}
We thank Brad Holden for prompt responses to many APF questions, the night operations team at Las Campanas Observatory for support of the MIKE observations, Ian Thompson for advice about MIKE and RVs, Dominick Rowan for help interpreting the lightcurve of HD 342092, Jhon Yana Galarza for help with MIKE calibrations and discussions related to RVs, Pranav Nagarajan for discussions on Nagarajan et al. 2025, and Natsuko Yamaguchi for sharing simulated data.
C.Y.L. acknowledges support from the Harrison and Carnegie Fellowships.

This work has made use of data from the European Space Agency (ESA) mission
{\it Gaia} (\url{https://www.cosmos.esa.int/gaia}), processed by the {\it Gaia}
Data Processing and Analysis Consortium (DPAC,
\url{https://www.cosmos.esa.int/web/gaia/dpac/consortium}). Funding for the DPAC
has been provided by national institutions, in particular the institutions
participating in the {\it Gaia} Multilateral Agreement.

This paper includes data gathered with the 6.5 meter Magellan Telescopes located at Las Campanas Observatory, Chile.

This work has made use of NASA’s Astrophysics Data System.

This research has made use of the SIMBAD database, operated at CDS, Strasbourg, France.

Guoshoujing Telescope (the Large Sky Area Multi-Object Fiber Spectroscopic Telescope LAMOST) is a National Major Scientific Project built by the Chinese Academy of Sciences. 
Funding for the project has been provided by the National Development and Reform Commission. 
LAMOST is operated and managed by the National Astronomical Observatories, Chinese Academy of Sciences.

Funding for SDSS-III has been provided by the Alfred P. Sloan Foundation, the Participating Institutions, the National Science Foundation, and the U.S. Department of Energy Office of Science. The SDSS-III web site is http://www.sdss3.org/.
SDSS-III is managed by the Astrophysical Research Consortium for the Participating Institutions of the SDSS-III Collaboration including the University of Arizona, the Brazilian Participation Group, Brookhaven National Laboratory, Carnegie Mellon University, University of Florida, the French Participation Group, the German Participation Group, Harvard University, the Instituto de Astrofisica de Canarias, the Michigan State/Notre Dame/JINA Participation Group, Johns Hopkins University, Lawrence Berkeley National Laboratory, Max Planck Institute for Astrophysics, Max Planck Institute for Extraterrestrial Physics, New Mexico State University, New York University, Ohio State University, Pennsylvania State University, University of Portsmouth, Princeton University, the Spanish Participation Group, University of Tokyo, University of Utah, Vanderbilt University, University of Virginia, University of Washington, and Yale University.

Funding for Rave has been provided by:
the Leibniz Institute for Astrophysics Potsdam (AIP);
the Australian Astronomical Observatory;
the Australian National University;
the Australian Research Council;
the French National Research Agency;
the German Research Foundation (SPP 1177 and SFB 881);
the European Research Council (ERC-StG 240271 Galactica);
the Istituto Nazionale di Astrofisica at Padova;
the Johns Hopkins University;
the National Science Foundation of the USA (AST-0908326);
the W. M. Keck foundation;
the Macquarie University;
the Netherlands Research School for Astronomy;
the Natural Sciences and Engineering Research Council of Canada;
the Slovenian Research Agency;
the Swiss National Science Foundation;
the Science \& Technology Facilities Council of the UK;
Opticon;
Strasbourg Observatory;
the Universities of Basel, Groningen, Heidelberg and Sydney.

Funding for the Sloan Digital Sky Survey IV has been provided by the Alfred P. Sloan Foundation, the U.S. Department of Energy Office of Science, and the Participating Institutions. SDSS acknowledges support and resources from the Center for High-Performance Computing at the University of Utah. The SDSS web site is www.sdss4.org.
SDSS is managed by the Astrophysical Research Consortium for the Participating Institutions of the SDSS Collaboration including the Brazilian Participation Group, the Carnegie Institution for Science, Carnegie Mellon University, Center for Astrophysics | Harvard \& Smithsonian (CfA), the Chilean Participation Group, the French Participation Group, Instituto de Astrofísica de Canarias, The Johns Hopkins University, Kavli Institute for the Physics and Mathematics of the Universe (IPMU) / University of Tokyo, the Korean Participation Group, Lawrence Berkeley National Laboratory, Leibniz Institut für Astrophysik Potsdam (AIP), Max-Planck-Institut für Astronomie (MPIA Heidelberg), Max-Planck-Institut für Astrophysik (MPA Garching), Max-Planck-Institut für Extraterrestrische Physik (MPE), National Astronomical Observatories of China, New Mexico State University, New York University, University of Notre Dame, Observatório Nacional / MCTI, The Ohio State University, Pennsylvania State University, Shanghai Astronomical Observatory, United Kingdom Participation Group, Universidad Nacional Autónoma de México, University of Arizona, University of Colorado Boulder, University of Oxford, University of Portsmouth, University of Utah, University of Virginia, University of Washington, University of Wisconsin, Vanderbilt University, and Yale University.

This paper analyzes data collected with the TESS mission, obtained from the MAST data archive at the Space Telescope Science Institute (STScI). 
Funding for the TESS mission is provided by the NASA Explorer Program. 
STScI is operated by the Association of Universities for Research in Astronomy, Inc., under NASA contract NAS 5–26555.

\end{acknowledgments}

\begin{contribution}

C.Y.L. initiated the project, conducted the MIKE observations, performed the analysis, and drafted the manuscript.
C.Y.L., J.D.S., K.E-B. assisted in planning the MIKE observations.
C.Y.L, J.D.S., K.E-B., J.L., and H.I. assisted in planning the APF observations.
J.D.S., K.E-B., H.I., and D.D.K. advised on the data reduction and RV analysis.
All authors read and provided feedback on the manuscript draft.

\end{contribution}

\facilities{Magellan:Clay (MIKE), APF, Gaia, Sloan, LAMOST, UKST}

\appendix

\section{Additional details on non-single star solutions \label{app:Additional details on non-single star solutions}}

\tabletypesize{\small}
\begin{deluxetable}{rc}
\tablecaption{Non-single stars}
\label{tab:Non-single stars}
\tablehead{\colhead{NSS solution} & 
    \colhead{Number}}
\startdata
\hline
Acceleration7 & 44 \\
Acceleration9 & 45 \\
FirstDegreeTrendSB1 & 8 \\
SecondDegreeTrendSB1 & 13 \\
Orbital & 71 \\
AstroSpectroSB1 & 67 \\
SB1 & 87 \\
EclipsingBinary & 1 \\
OrbitalTargetedSearch & 1 \\
\enddata
% \tablecomments{}
\end{deluxetable}

Table \ref{tab:Non-single stars} provides the types and number of non-single star solutions in our sample.
The specific model parameterizations are described in \cite{Arenou:2023}, \cite{Halbwachs:2023} (astrometry), and \cite{Gosset:2025} (spectroscopy).
There are a total of 337 solutions, which is more than the 304 solutions mentioned in Section \ref{sec:Sample selection}; this is because 
33 sources have two solutions; we describe these in more detail below.

8 of the sources have SB1 and Orbital solutions that were not combined. 
However, the two different orbital periods were consistent with each other to within $3\sigma$ in all cases, and the eccentricities were consistent with each other to within $3\sigma$ in all but one case.
This suggests that the SB1 and Orbital solutions are likely tracing the same orbit, but there was some issue in the matching procedure that resulted in the solutions remaining uncombined.
7 of these sources have fairly long orbital periods ($\approx 600 - 1200$ days), 1 source has a shorter orbital period of $\approx 230$ days.
% 1050229047014245120, 1109631635877569024, 1474455370705908224, 1758531826288206080, 1842966313482671616, 2164592289448420224, 3552582374981956736, 4382114042159988352

9 of the sources have an astrometric or spectroscopic orbit with a fairly long $500 - 1400$ day orbital period, and the complementary astrometric or spectroscopic acceleration solution.
7 of the sources have both an astrometric and spectroscopic acceleration solution.
% 6903293654893289856, 6855269662754949248, 6451215048105801216, 6323634796925577216, 6303678390298802944, 5874533710639696000, 4382040817260949376
It is likely that all these sources have genuinely long orbital periods.

5 of the sources have SB1 and Orbital solutions that were not combined, where the SB1 solution has a short orbital period $\approx 1 - 30$ days but the Orbital solution is $\approx 300 - 600$ days; this suggests a hierarchical triple, with a tight inner (spectroscopic) binary and a wider outer (astrometric) tertiary component.
2 sources have similarly short orbital periods and an astrometric acceleration solution, suggesting the same explanation.
2 sources have $100 - 200$ day spectroscopic orbits, but astrometric acceleration solutions; the interpretation here is less clear (e.g., a very unlucky lack of an orbital solution, or a very wide astrometric tertiary).
These 9 sources are listed in Table \ref{tab:Potential hierarchical triples} for reference.

\tabletypesize{\small}
\begin{deluxetable}{ccc}
\tablecaption{Potential hierarchical triples}
\label{tab:Potential hierarchical triples}
\tablehead{\colhead{Gaia DR3 source ID} & 
    \colhead{Spec. $P_{orb}$ (day)} &
    \colhead{Ast. $P_{orb}$ (day)}}
\startdata
\hline
2000491037995373184 & 0.60 $\pm$ 0.00 & 316.90 $\pm$ 0.86 \\
4232766140808846464 & 1.70 $\pm$ 0.00 & 703.94 $\pm$ 28.60 \\
4498980102287082624 & 4.30 $\pm$ 0.00 & 568.36 $\pm$ 1.57 \\
5240296955165692032 & 13.54 $\pm$ 0.01 & 447.28 $\pm$ 2.14 \\
6369935265930164608 & 1.37 $\pm$ 0.00 & 546.67 $\pm$ 6.27 \\
\hline
3714004421538539904 & 2.52 $\pm$ 0.00 & Acceleration9 \\
1256697745659210368 & 30.73 $\pm$ 0.02 & Acceleration7 \\
\hline
5200175324859556352 & 113.99 $\pm$ 0.25 & Acceleration9 \\
1911398607229097728 & 211.32 $\pm$ 3.46 & Acceleration7 \\
\enddata
% \tablecomments{}
\end{deluxetable}

\section{Archival data \label{app:Archival data}}
% https://www.lamost.org/dr10/v1.0/medcas/search
% https://www.lamost.org/dr10/v1.0/search

We cross-match our sample with the Large Sky Area Multi-Object Fiber Spectroscopic Telescope \citep[LAMOST; ][]{Cui:2012} Data Release 10 v1.0, which contains spectra taken between 2011 October 24 through 2022 June 30.
% https://www.lamost.org/dr11/v1.0/doc/mr-data-production-description
% https://www.lamost.org/dr11/v1.0/doc/lr-data-production-description
The low resolution spectrograph has resolving power $R\sim1500$ over 3700--9000\AA\, with the blue (red) arm covering wavelengths $<5900$\AA\, ($>5700$)\AA. 
The medium resolution spectrograph has $R\sim7500$ over 4950 -- 5350\AA\, (blue arm) and 6300 -- 6800\AA\, (red arm).
LAMOST provides coverage in the northern hemisphere.

We also search for archival spectra from the Sloan Digital Sky Survey (SDSS) Apache Point Observatory Galactic Evolution Experiment \citep[APOGEE;][]{Majewski:2017}.
The APOGEE spectrographs have resolving power $R \sim 22,500$ covering the NIR $1.51–1.70 \mu$m wavelength range.
We search both the SDSS-III \citep{Alam:2015} and SDSS-IV \citep{Abdurro'uf:2022} final data releases.
APOGEE provides coverage in both the northern and southern hemispheres.

Finally, we search the Radial Velocity Experiment (RAVE) survey's Data Release 6 \citep{Steinmetz:2020}, which contains spectra taken between 2003 April 12 and 2013 April 4.
RAVE produces medium-resolution spectra $R \sim 7500$ covering the Ca II triplet region 8410--8795 \AA. 
RAVE provides coverage in the southern hemisphere.

For the SDSS sources, we propagate the positions to J2000.0 using the Gaia proper motions, and define a cross-match to be the closest source within 3\arcsec.
LAMOST and RAVE are already cross-matched to Gaia DR3.
We impose a quality cut on the RVs, requiring the RV error to be less than 5 km s$^{-1}$ for the LAMOST low-resolution spectra, and less than 3 km s$^{-1}$ for all the other spectra.
The statistics on the number of spectra available are listed in Table \ref{tab:Archival spectra}.
218 unique sources have archival spectra; 85 are NSS sources (54 with orbital solutions) and 133 are only in the single-star catalog.

\tabletypesize{\small}
\begin{deluxetable}{rcc}
\tablecaption{Archival spectra}
\label{tab:Archival spectra}
\tablehead{\colhead{Survey} & 
    \colhead{Number of spectra} & 
    \colhead{Number of targets}}
\startdata
\hline
RAVE & 57 & 47 \\
LAMOST MRS & 326 & 80 \\ 
LAMOST LRS & 164 & 104 \\ 
APOGEE & 243 & 47 \\
\hline
Total & 790 & 218 \\
\enddata
\tablecomments{The total number of targets is not the sum of the above columns, because some sources are in multiple surveys.}
\end{deluxetable}

For those where we have the APF or MIKE spectrum of the source and determined its Gaia RV to be unreliable (Section \ref{sec:Removal of unreliable Gaia RVs}), we do not analyze the archival RV.

We look for any archival RVs that meet our criteria in Section \ref{sec:Interesting targets from radial velocity follow-up}, i.e., $\drv > 60$ km s$^{-1}$ or $|RV_{APF/MIKE} - \rvgaia| > 15$ km s$^{-1}$ and $\drv > 30$ km s$^{-1}$.
We do not place any constraint on when the observation was taken.
We identify 14 sources that have archival data that meet these criteria.

7 of these sources are known false positives; 5 are identified elsewhere in this paper. 
The other two that have not been discussed are Gaia DR3 660482789494668928 (not yet observed, but a known eclipsing binary named MU Cnc), and Gaia DR3 3112344688094507136 (has been observed, is a binary post-AGB named SZ Mon). 
The 7 remaining sources are plausible candidates, which we discuss here.

Gaia DR3 3879659596453350400 has one LAMOST LRS observation on 2016-11-27 and one LAMOST MRS observation on 2022-04-07, which are 27 km s$^{-1}$ and 36 km s$^{-1}$ below $\rvgaia$, respectively.
The LAMOST RVs are roughly consistent with the $\drv = 59$ km s$^{-1}$, assuming the amplitude is centered on $\rvgaia = 44$ km s$^{-1}$.
Gaia DR3 3879659596453350400 has an astrometric orbital solution ($P_{orb} = 308.5$ day, $\alpha_0 = 0.87$ mas); it does not have a spectroscopic solution. 
Given its parallax of $\varpi = 3.08$ mas, this corresponds to an astrometric mass function of $0.03 M_\odot$. 
It has a goodness-of-fit = 7.41 and significance = 28.93, so it is a plausible solution.
It is unclear why it is missing a spectroscopic solution, except that the Ca II triplet is weak which may have made the velocities more unreliable.
The astrometric fit prefers an edge-on orbit, which would also be consistent with a stellar binary explanation at this RV amplitude.
We have not observed this source because it has an orbital solution.
We regard this candidate as unlikely to be a BH.
% Gaia RV: 44.12 +/- 5.77 km/s, amplitude = 58.77 km/s
% LAMOST : LRS 2016-11-27, 16.74 km/s; MRS 2022-04-07, 7.63

Gaia DR3 1030880524287375360 has 5 epochs of LAMOST MRS RV (observations taken 2021-01-01, 2022-01-12, 2022-01-13, 2022-01-15, 2022-01-17) within $\pm 3$ km s$^{-1}$ of $\rvgaia$ and which are consistent given the measurement uncertainties. 
There is 1 epoch of LAMOST LRS RV observations from 2014-02-10 which is 6 km s$^{-1}$ lower than $\rvgaia$ but which is consistent within $1.5\sigma$.
There is a single discrepant LAMOST MRS RV observation on 2022-02-15 that is 41 km s$^{-1}$ lower than the Gaia RV.
This would put it in tension with the $\drv = 41$ km s$^{-1}$, assuming it is centered on $\rvgaia$.
We inspect the LAMOST spectrum of this source, and find that the line depth and shape of the Mg b triplet and H$\alpha$ absorption lines differ from the other MRS RV epochs.
We thus regard this discrepant RV epoch and its corresponding RV as unreliable.
We observed this source on 2024-10-02 with APF and measured an RV 6 km s$^{-1}$ lower than $\rvgaia$.
We regard this candidate as unlikely to actually have met our intended selection criteria for large RV difference from $\rvgaia$.
% LAMOST: RV = 
% Gaia: RV = 32.16 +/- 2.11 km/s, amplitude = 40.93 km/s
% Me: 2024-10-02, 26.42 km/s

Gaia DR3 1115628342232842368 has a LAMOST MRS RV observation on 2018-11-15 which is 19 km s$^{-1}$ above $\rvgaia$.
This is an astrometrically accelerating source in our sample that we have not yet observed.
The LAMOST RV is consistent with the $\drv = 38$ km s$^{-1}$, assuming the amplitude is centered on $\rvgaia$.
% LAMOST: 2018-11-15, RV = -316.71 km/s
% Gaia: RV = -335.43 +/- 2.27 km/s, amplitude = 38.16 km/s

Gaia DR3 1590423439067313664 has a LAMOST LRS RV observation on 2013-04-21 that is 16 km s$^{-1}$ higher than $\rvgaia$.
However, the uncertainty on $\rvgaia$ itself is 35 km s$^{-1}$, which means it is consistent with $\rvgaia$; this is possibly because the Ca II triplet is weak.
Nonetheless, it has an AstroSpectroSB1 orbit ($P_{orb} = 326.4$ day, $\alpha_0 = 1.16$ mas, significance = 92, goodness-of-fit = 4). 
Given its parallax of $\varpi = 3.21$ mas, it has an astrometric mass function of $0.06 M_\odot$.
It is in the Silver sample of \cite{Andrew:2022}.
We did not observe this source because it has an orbital solution.
% LAMOST: 2013-04-21, RV = -37.98 km/s
% Gaia: RV = -54.23 +/- 34.83 km/s, amplitude = 40.66 km/s.
% Not variable.
% Also a case with RV error larger than RV amplitude robust, and very large.

Gaia DR3 3353117427967936768 has a LAMOST LRS RV observation on 2017-10-15 which is 16 km s$^{-1}$ smaller than $\rvgaia$.
We observed this source with APF on 2024-12-02, and measured an RV consistent with the LAMOST RV, within the uncertainties.
The LAMOST RV is within the $\drv = 36$ km s$^{-1}$, assuming the amplitude is centered on $\rvgaia$.
$\rvgaia$ itself is relatively uncertain ($\pm 8$ km s$^{-1}$), possibly because the Ca II triplet is somewhat weak.
This source is not in any of the NSS catalogs, but has a large $\ruwe = 16.70$ and relatively small parallax $\varpi = 0.61 \pm 0.26$ mas.
Assuming this is a binary and not a hierarchical triple, this would suggest this source might have a long orbital period.
% LAMOST: MJD = 58041 (MM-DD-YYYY 10-15-2017), RV = -7.99 km/s
% Gaia: RV = 8.29 +/- 7.61 km/s, amplitude = 36.48 km/s.
% Me: 2024-12-02, RV = -6.25 km/s
% Not variable.

Gaia DR3 3850375989737564160 has one LAMOST MRS RV observation on 2021-12-19 which is 16 km s$^{-1}$ above $\rvgaia$.
It also has two LAMOST LRS RV observations on 2015-04-20 and 2018-20-18, which are 10 km s$^{-1}$ below and 10 km s$^{-1}$ above $\rvgaia$, respectively).
We observed this source on 2024-12-19 with MIKE, and our RV is less than 2 km s$^{-1}$ different than $\rvgaia$, and within the uncertainties, which is why it was not included in Table \ref{tab:rv_amp_30_drv_15}.
All LAMOST RVs are well within the $\drv = 58$ km s$^{-1}$, assuming the amplitude is centered on $\rvgaia$.
This source is not in any of the NSS catalogs, but has a large $\ruwe = 7.66$ and relatively small parallax ($\varpi = 0.62 \pm 0.11$ mas).
We draw a similar conclusion to Gaia DR3 3353117427967936768 regarding the possible orbital period of this source.
% LAMOST: MJD = 59567 (MM-DD-YYYY 12-19-2021), RV = -32.86 km/s
% Gaia: RV = -48.85 km/s, amplitude = 57.74 km/s.
% Me: 2024-12-19, RV = -50.30 km/s
% Not variable.

Gaia DR3 6196341900531496832 has been identified in the main text. 

\section{Additional cross-matches to compact object candidate catalogs and searches \label{app:Additional cross-matches to compact object candidate catalogs and searches}}

Below are brief description of several catalogs not mentioned in Section \ref{sec:Comparison to other compilations of compact object candidates}.
All samples were selected from the Gaia DR3 non-single star catalogs unless otherwise mentioned.

Our sample does not have any overlap with the 234 candidates in \cite{Jayasinghe:2023} selected from SB1 solutions with high-mass function ($f_{M,S} > 1.5 M_\odot$ for evolved stars, $f_{M,S} > 0.5 M_\odot$ for main-sequence sources).
\cite{Jayasinghe:2023} determined the majority of the sources were false positives such as eclipsing binaries, ellipsoidal variables, luminous star binaries, and did not identify any strong compact object candidates. 

Our sample does not have any overlap with the 14 candidates in \cite{El-Badry:2022} and Table 9 of \cite{Arenou:2023} which were selected from SB1 solutions to have inferred secondary masses greater than $3 M_\odot$ and $K/\sigma_K > 20$.
\cite{El-Badry:2022} found that all the sources were composed of mass-transfer binaries with a highly stripped low-mass giant donor and an intermediate-mass main-sequence accretor.

Our sample does not have any overlap with the 24 candidates in \cite{Andrews:2022}, which were selected to have large astrometric mass functions assuming that all the luminous sources were $1 M_\odot$ and the companions were dark, small goodness-of-fits, and non-giant companions.
Although no BHs have been found in their list (Gaia BH1 was excluded due to suspicion over its $\approx 0.5$ year orbital period), several white dwarf and/or neutron star candidates have been subsequently identified \citep{Geier:2023, El-Badry:2024d, El-Badry:2024b}.

Our sample does not have any overlap with the 6 candidates in \cite{El-Badry:2023a}, which were selected to have astrometric mass functions greater than $3M_\odot$ and large $a_0$ at a given $P_{orb}$.
This sample included Gaia BH1 and BH2.

Our sample does not have any overlap with the the 10 candidates in \cite{Chakrabarti:2023} which were selected to have secondary masses $M_2 > 5M_\odot$ and did not show signs of having a luminous companion such as sources classified as SB2s or having eclipses.
This sample included Gaia BH1.

Our sample does not have any overlap with the 34 candidates in Simon et al. in prep.
The candidates in Simon et al. in prep were selected from both the orbit and acceleration catalogs. 
The orbital candidates were selected to have secondary mass estimates at least $5M_\odot$.
The acceleration candidates were selected to have large accelerations both along and transverse to the line of sight, i.e., both astrometric and spectroscopic accelerations.

14 of our candidates were in the catalog of \cite{Andrew:2022}.
1 was classified as gold (Gaia DR3 4441393313920391936),
12 were classified as silver (Gaia DR3 1052266579499163392, 1546566669018411264, 1650922176599904000, 1940380874743387776, 4458199593966253184, 4852879978634032128, 5511004204015081472, 6408548258871775104, 6525834474174171264, 289423281843928448, 1590423439067313664, 5721114077155913984), and 1 was classified as bronze (6845526207324092032).
\cite{Andrew:2022} selected compact object candidates by identifying sources that had large astrometric and spectroscopic deviations from single-star behavior.
They then classified their candidates as gold, silver, or bronze candidates, with the gold and silver candidates having additional selection criteria.
The candidates in \cite{Andrew:2022} contains hierarchical stellar triples and compact objects; gold classified sources are most likely to have compact objects but the sample is still likely contaminated.

\bibliography{sample701}{}

\begin{thebibliography}{}
\expandafter\ifx\csname natexlab\endcsname\relax\def\natexlab#1{#1}\fi
\providecommand{\url}[1]{\href{#1}{#1}}
\providecommand{\dodoi}[1]{doi:~\href{http://doi.org/#1}{\nolinkurl{#1}}}
\providecommand{\doeprint}[1]{\href{http://ascl.net/#1}{\nolinkurl{http://ascl.net/#1}}}
\providecommand{\doarXiv}[1]{\href{https://arxiv.org/abs/#1}{\nolinkurl{https://arxiv.org/abs/#1}}}

% type= article
\bibitem[{R. {Abbott} {et~al.}(2020{\natexlab{a}}){Abbott}, {Abbott}, {Abraham}, {Acernese}, {Ackley}, {Adams}, {Adhikari}, {Adya}, {Affeldt}, {Agathos}, {Agatsuma}, {Aggarwal}, {Aguiar}, {Aich}, {Aiello}, {Ain}, {Ajith}, {Akcay}, {Allen}, {Allocca}, {Altin}, {Amato}, {Anand}, {Ananyeva}, {Anderson}, {Anderson}, {Angelova}, {Ansoldi}, {Antier}, {Appert}, {Arai}, {Araya}, {Areeda}, {Ar{\`e}ne}, {Arnaud}, {Aronson}, {Arun}, {Asali}, {Ascenzi}, {Ashton}, {Aston}, {Astone}, {Aubin}, {Aufmuth}, {AultONeal}, {Austin}, {Avendano}, {Babak}, {Bacon}, {Badaracco}, {Bader}, {Bae}, {Baer}, {Baird}, {Baldaccini}, {Ballardin}, {Ballmer}, {Bals}, {Balsamo}, {Baltus}, {Banagiri}, {Bankar}, {Bankar}, {Barayoga}, {Barbieri}, {Barish}, {Barker}, {Barkett}, {Barneo}, {Barone}, {Barr}, {Barsotti}, {Barsuglia}, {Barta}, {Bartlett}, {Bartos}, {Bassiri}, {Basti}, {Bawaj}, {Bayley}, {Bazzan}, {B{\'e}csy}, {Bejger}, {Belahcene}, {Bell}, {Beniwal}, {Benjamin}, {Bentley}, {Bergamin}, {Berger}, {Bergmann}, {Bernuzzi}, {Berry},
  {Bersanetti}, {Bertolini}, {Betzwieser}, {Bhandare}, {Bhandari}, {Bidler}, {Biggs}, {Bilenko}, {Billingsley}, {Birney}, {Birnholtz}, {Biscans}, {Bischi}, {Biscoveanu}, {Bisht}, {Bissenbayeva}, {Bitossi}, {Bizouard}, {Blackburn}, {Blackman}, {Blair}, {Blair}, {Blair}, {Bobba}, {Bode}, {Boer}, {Boetzel}, {Bogaert}, {Bondu}, {Bonilla}, {Bonnand}, {Booker}, {Boom}, {Bork}, {Boschi}, {Bose}, {Bossilkov}, {Bosveld}, {Bouffanais}, {Bozzi}, {Bradaschia}, {Brady}, {Bramley}, {Branchesi}, {Brau}, {Breschi}, {Briant}, {Briggs}, {Brighenti}, {Brillet}, {Brinkmann}, {Brockill}, {Brooks}, {Brooks}, {Brown}, {Brunett}, {Bruno}, {Bruntz}, {Buikema}, {Bulik}, {Bulten}, {Buonanno}, {Buscicchio}, {Buskulic}, {Byer}, {Cabero}, {Cadonati}, {Cagnoli}, {Cahillane}, {Bustillo}, {Callaghan}, {Callister}, {Calloni}, {Camp}, {Canepa}, {Cannon}, {Cao}, {Cao}, {Carapella}, {Carbognani}, {Caride}, {Carney}, {Carullo}, {Diaz}, {Casentini}, {Casta{\~n}eda}, {Caudill}, {Cavagli{\`a}}, {Cavalier}, {Cavalieri}, {Cella},
  {Cerd{\'a}-Dur{\'a}n}, {Cesarini}, {Chaibi}, {Chakravarti}, {Chan}, {Chan}, {Chao}, {Charlton}, {Chase}, {Chassande-Mottin}, {Chatterjee}, {Chaturvedi}, {Chatziioannou}, {Chen}, {Chen}, \& {Chen}}]{Abbott_GW190521:2020}
{Abbott}, R., {Abbott}, T.~D., {Abraham}, S., {et~al.} 2020{\natexlab{a}}, \bibinfo{title}{{Properties and Astrophysical Implications of the 150 M$_{{\ensuremath{\odot}}}$ Binary Black Hole Merger GW190521},} \apjl, 900, L13, \dodoi{10.3847/2041-8213/aba493}

% type= article
\bibitem[{R. {Abbott} {et~al.}(2020{\natexlab{b}}){Abbott}, {Abbott}, {Abraham}, {Acernese}, {Ackley}, {Adams}, {Adhikari}, {Adya}, {Affeldt}, {Agathos}, {Agatsuma}, {Aggarwal}, {Aguiar}, {Aich}, {Aiello}, {Ain}, {Ajith}, {Akcay}, {Allen}, {Allocca}, {Altin}, {Amato}, {Anand}, {Ananyeva}, {Anderson}, {Anderson}, {Angelova}, {Ansoldi}, {Antier}, {Appert}, {Arai}, {Araya}, {Areeda}, {Ar{\`e}ne}, {Arnaud}, {Aronson}, {Arun}, {Asali}, {Ascenzi}, {Ashton}, {Aston}, {Astone}, {Aubin}, {Aufmuth}, {AultONeal}, {Austin}, {Avendano}, {Babak}, {Bacon}, {Badaracco}, {Bader}, {Bae}, {Baer}, {Baird}, {Baldaccini}, {Ballardin}, {Ballmer}, {Bals}, {Balsamo}, {Baltus}, {Banagiri}, {Bankar}, {Bankar}, {Barayoga}, {Barbieri}, {Barish}, {Barker}, {Barkett}, {Barneo}, {Barone}, {Barr}, {Barsotti}, {Barsuglia}, {Barta}, {Bartlett}, {Bartos}, {Bassiri}, {Basti}, {Bawaj}, {Bayley}, {Bazzan}, {B{\'e}csy}, {Bejger}, {Belahcene}, {Bell}, {Beniwal}, {Benjamin}, {Benkel}, {Bentley}, {Bergamin}, {Berger}, {Bergmann}, {Bernuzzi}, {Berry},
  {Bersanetti}, {Bertolini}, {Betzwieser}, {Bhandare}, {Bhandari}, {Bidler}, {Biggs}, {Bilenko}, {Billingsley}, {Birney}, {Birnholtz}, {Biscans}, {Bischi}, {Biscoveanu}, {Bisht}, {Bissenbayeva}, {Bitossi}, {Bizouard}, {Blackburn}, {Blackman}, {Blair}, {Blair}, {Blair}, {Bobba}, {Bode}, {Boer}, {Boetzel}, {Bogaert}, {Bondu}, {Bonilla}, {Bonnand}, {Booker}, {Boom}, {Bork}, {Boschi}, {Bose}, {Bossilkov}, {Bosveld}, {Bouffanais}, {Bozzi}, {Bradaschia}, {Brady}, {Bramley}, {Branchesi}, {Brau}, {Breschi}, {Briant}, {Briggs}, {Brighenti}, {Brillet}, {Brinkmann}, {Brito}, {Brockill}, {Brooks}, {Brooks}, {Brown}, {Brunett}, {Bruno}, {Bruntz}, {Buikema}, {Bulik}, {Bulten}, {Buonanno}, {Buskulic}, {Byer}, {Cabero}, {Cadonati}, {Cagnoli}, {Cahillane}, {Bustillo}, {Callaghan}, {Callister}, {Calloni}, {Camp}, {Canepa}, {Cannon}, {Cao}, {Cao}, {Carapella}, {Carbognani}, {Caride}, {Carney}, {Carullo}, {Diaz}, {Casentini}, {Casta{\~n}eda}, {Caudill}, {Cavagli{\`a}}, {Cavalier}, {Cavalieri}, {Cella}, {Cerd{\'a}-Dur{\'a}n},
  {Cesarini}, {Chaibi}, {Chakravarti}, {Chan}, {Chan}, {Chao}, {Charlton}, {Chase}, {Chassande-Mottin}, {Chatterjee}, {Chaturvedi}, {Chatziioannou}, {Chen}, \& {Chen}}]{Abbott_GW190814:2020}
{Abbott}, R., {Abbott}, T.~D., {Abraham}, S., {et~al.} 2020{\natexlab{b}}, \bibinfo{title}{{GW190814: Gravitational Waves from the Coalescence of a 23 Solar Mass Black Hole with a 2.6 Solar Mass Compact Object},} \apjl, 896, L44, \dodoi{10.3847/2041-8213/ab960f}

% type= article
\bibitem[{ {Abdurro'uf} {et~al.}(2022){Abdurro'uf}, {Accetta}, {Aerts}, {Silva Aguirre}, {Ahumada}, {Ajgaonkar}, {Filiz Ak}, {Alam}, {Allende Prieto}, {Almeida}, {Anders}, {Anderson}, {Andrews}, {Anguiano}, {Aquino-Ort{\'\i}z}, {Arag{\'o}n-Salamanca}, {Argudo-Fern{\'a}ndez}, {Ata}, {Aubert}, {Avila-Reese}, {Badenes}, {Barb{\'a}}, {Barger}, {Barrera-Ballesteros}, {Beaton}, {Beers}, {Belfiore}, {Bender}, {Bernardi}, {Bershady}, {Beutler}, {Bidin}, {Bird}, {Bizyaev}, {Blanc}, {Blanton}, {Boardman}, {Bolton}, {Boquien}, {Borissova}, {Bovy}, {Brandt}, {Brown}, {Brownstein}, {Brusa}, {Buchner}, {Bundy}, {Burchett}, {Bureau}, {Burgasser}, {Cabang}, {Campbell}, {Cappellari}, {Carlberg}, {Wanderley}, {Carrera}, {Cash}, {Chen}, {Chen}, {Cherinka}, {Chiappini}, {Choi}, {Chojnowski}, {Chung}, {Clerc}, {Cohen}, {Comerford}, {Comparat}, {da Costa}, {Covey}, {Crane}, {Cruz-Gonzalez}, {Culhane}, {Cunha}, {Dai}, {Damke}, {Darling}, {Davidson}, {Davies}, {Dawson}, {De Lee}, {Diamond-Stanic}, {Cano-D{\'\i}az}, {S{\'a}nchez},
  {Donor}, {Duckworth}, {Dwelly}, {Eisenstein}, {Elsworth}, {Emsellem}, {Eracleous}, {Escoffier}, {Fan}, {Farr}, {Feng}, {Fern{\'a}ndez-Trincado}, {Feuillet}, {Filipp}, {Fillingham}, {Frinchaboy}, {Fromenteau}, {Galbany}, {Garc{\'\i}a}, {Garc{\'\i}a-Hern{\'a}ndez}, {Ge}, {Geisler}, {Gelfand}, {G{\'e}ron}, {Gibson}, {Goddy}, {Godoy-Rivera}, {Grabowski}, {Green}, {Greener}, {Grier}, {Griffith}, {Guo}, {Guy}, {Hadjara}, {Harding}, {Hasselquist}, {Hayes}, {Hearty}, {Hern{\'a}ndez}, {Hill}, {Hogg}, {Holtzman}, {Horta}, {Hsieh}, {Hsu}, {Hsu}, {Huber}, {Huertas-Company}, {Hutchinson}, {Hwang}, {Ibarra-Medel}, {Chitham}, {Ilha}, {Imig}, {Jaekle}, {Jayasinghe}, {Ji}, {Johnson}, {Jones}, {J{\"o}nsson}, {Katkov}, {Khalatyan}, {Kinemuchi}, {Kisku}, {Knapen}, {Kneib}, {Kollmeier}, {Kong}, {Kounkel}, {Kreckel}, {Krishnarao}, {Lacerna}, {Lane}, {Langgin}, {Lavender}, {Law}, {Lazarz}, {Leung}, {Leung}, {Lewis}, {Li}, {Li}, {Lian}, {Liang}, {Lin}, {Lin}, {Lin}, {Lintott}, {Long}, {Longa-Pe{\~n}a}, {L{\'o}pez-Cob{\'a}}, {Lu},
  {Lundgren}, {Luo}, {Mackereth}, {de la Macorra}, {Mahadevan}, {Majewski}, {Manchado}, {Mandeville}, {Maraston}, {Margalef-Bentabol}, {Masseron}, {Masters}, {Mathur}, {McDermid}, {Mckay}, {Merloni}, {Merrifield}, {Meszaros}, {Miglio}, {Di Mille}, {Minniti}, {Minsley}, \& {Monachesi}}]{Abdurro'uf:2022}
{Abdurro'uf}, {Accetta}, K., {Aerts}, C., {et~al.} 2022, \bibinfo{title}{{The Seventeenth Data Release of the Sloan Digital Sky Surveys: Complete Release of MaNGA, MaStar, and APOGEE-2 Data},} \apjs, 259, 35, \dodoi{10.3847/1538-4365/ac4414}

% type= article
\bibitem[{S. {Alam} {et~al.}(2015){Alam}, {Albareti}, {Allende Prieto}, {Anders}, {Anderson}, {Anderton}, {Andrews}, {Armengaud}, {Aubourg}, {Bailey}, {Basu}, {Bautista}, {Beaton}, {Beers}, {Bender}, {Berlind}, {Beutler}, {Bhardwaj}, {Bird}, {Bizyaev}, {Blake}, {Blanton}, {Blomqvist}, {Bochanski}, {Bolton}, {Bovy}, {Shelden Bradley}, {Brandt}, {Brauer}, {Brinkmann}, {Brown}, {Brownstein}, {Burden}, {Burtin}, {Busca}, {Cai}, {Capozzi}, {Carnero Rosell}, {Carr}, {Carrera}, {Chambers}, {Chaplin}, {Chen}, {Chiappini}, {Chojnowski}, {Chuang}, {Clerc}, {Comparat}, {Covey}, {Croft}, {Cuesta}, {Cunha}, {da Costa}, {Da Rio}, {Davenport}, {Dawson}, {De Lee}, {Delubac}, {Deshpande}, {Dhital}, {Dutra-Ferreira}, {Dwelly}, {Ealet}, {Ebelke}, {Edmondson}, {Eisenstein}, {Ellsworth}, {Elsworth}, {Epstein}, {Eracleous}, {Escoffier}, {Esposito}, {Evans}, {Fan}, {Fern{\'a}ndez-Alvar}, {Feuillet}, {Filiz Ak}, {Finley}, {Finoguenov}, {Flaherty}, {Fleming}, {Font-Ribera}, {Foster}, {Frinchaboy}, {Galbraith-Frew}, {Garc{\'\i}a},
  {Garc{\'\i}a-Hern{\'a}ndez}, {Garc{\'\i}a P{\'e}rez}, {Gaulme}, {Ge}, {G{\'e}nova-Santos}, {Georgakakis}, {Ghezzi}, {Gillespie}, {Girardi}, {Goddard}, {Gontcho}, {Gonz{\'a}lez Hern{\'a}ndez}, {Grebel}, {Green}, {Grieb}, {Grieves}, {Gunn}, {Guo}, {Harding}, {Hasselquist}, {Hawley}, {Hayden}, {Hearty}, {Hekker}, {Ho}, {Hogg}, {Holley-Bockelmann}, {Holtzman}, {Honscheid}, {Huber}, {Huehnerhoff}, {Ivans}, {Jiang}, {Johnson}, {Kinemuchi}, {Kirkby}, {Kitaura}, {Klaene}, {Knapp}, {Kneib}, {Koenig}, {Lam}, {Lan}, {Lang}, {Laurent}, {Le Goff}, {Leauthaud}, {Lee}, {Lee}, {Licquia}, {Liu}, {Long}, {L{\'o}pez-Corredoira}, {Lorenzo-Oliveira}, {Lucatello}, {Lundgren}, {Lupton}, {Mack}, {Mahadevan}, {Maia}, {Majewski}, {Malanushenko}, {Malanushenko}, {Manchado}, {Manera}, {Mao}, {Maraston}, {Marchwinski}, {Margala}, {Martell}, {Martig}, {Masters}, {Mathur}, {McBride}, {McGehee}, {McGreer}, {McMahon}, {M{\'e}nard}, {Menzel}, {Merloni}, {M{\'e}sz{\'a}ros}, {Miller}, {Miralda-Escud{\'e}}, {Miyatake}, {Montero-Dorta}, {More},
  {Morganson}, {Morice-Atkinson}, {Morrison}, {Mosser}, {Muna}, {Myers}, {Nandra}, {Newman}, {Neyrinck}, {Nguyen}, {Nichol}, {Nidever}, {Noterdaeme}, {Nuza}, {O'Connell}, {O'Connell}, {O'Connell}, {Ogando}, {Olmstead}, {Oravetz}, {Oravetz}, {Osumi}, {Owen}, {Padgett}, {Padmanabhan}, {Paegert}, {Palanque-Delabrouille}, \& {Pan}}]{Alam:2015}
{Alam}, S., {Albareti}, F.~D., {Allende Prieto}, C., {et~al.} 2015, \bibinfo{title}{{The Eleventh and Twelfth Data Releases of the Sloan Digital Sky Survey: Final Data from SDSS-III},} \apjs, 219, 12, \dodoi{10.1088/0067-0049/219/1/12}

% type= article
\bibitem[{R. {Andrae} {et~al.}(2023{\natexlab{a}}){Andrae}, {Rix}, \& {Chandra}}]{Andrae:2023b}
{Andrae}, R., {Rix}, H.-W., \& {Chandra}, V. 2023{\natexlab{a}}, \bibinfo{title}{{Robust Data-driven Metallicities for 175 Million Stars from Gaia XP Spectra},} \apjs, 267, 8, \dodoi{10.3847/1538-4365/acd53e}

% type= article
\bibitem[{R. {Andrae} {et~al.}(2023{\natexlab{b}}){Andrae}, {Fouesneau}, {Sordo}, {Bailer-Jones}, {Dharmawardena}, {Rybizki}, {De Angeli}, {Lindstr{\o}m}, {Marshall}, {Drimmel}, {Korn}, {Soubiran}, {Brouillet}, {Casamiquela}, {Rix}, {Abreu Aramburu}, {{\'A}lvarez}, {Bakker}, {Bellas-Velidis}, {Bijaoui}, {Brugaletta}, {Burlacu}, {Carballo}, {Chaoul}, {Chiavassa}, {Contursi}, {Cooper}, {Creevey}, {Dafonte}, {Dapergolas}, {de Laverny}, {Delchambre}, {Demouchy}, {Edvardsson}, {Fr{\'e}mat}, {Garabato}, {Garc{\'\i}a-Lario}, {Garc{\'\i}a-Torres}, {Gavel}, {Gomez}, {Gonz{\'a}lez-Santamar{\'\i}a}, {Hatzidimitriou}, {Heiter}, {Jean-Antoine Piccolo}, {Kontizas}, {Kordopatis}, {Lanzafame}, {Lebreton}, {Licata}, {Livanou}, {Lobel}, {Lorca}, {Magdaleno Romeo}, {Manteiga}, {Marocco}, {Mary}, {Nicolas}, {Ordenovic}, {Pailler}, {Palicio}, {Pallas-Quintela}, {Panem}, {Pichon}, {Poggio}, {Recio-Blanco}, {Riclet}, {Robin}, {Santove{\~n}a}, {Sarro}, {Schultheis}, {Segol}, {Silvelo}, {Slezak}, {Smart}, {S{\"u}veges},
  {Th{\'e}venin}, {Torralba Elipe}, {Ulla}, {Utrilla}, {Vallenari}, {van Dillen}, {Zhao}, \& {Zorec}}]{Andrae:2023a}
{Andrae}, R., {Fouesneau}, M., {Sordo}, R., {et~al.} 2023{\natexlab{b}}, \bibinfo{title}{{Gaia Data Release 3. Analysis of the Gaia BP/RP spectra using the General Stellar Parameterizer from Photometry},} \aap, 674, A27, \dodoi{10.1051/0004-6361/202243462}

% type= article
\bibitem[{S. {Andrew} {et~al.}(2022){Andrew}, {Penoyre}, {Belokurov}, {Evans}, \& {Oh}}]{Andrew:2022}
{Andrew}, S., {Penoyre}, Z., {Belokurov}, V., {Evans}, N.~W., \& {Oh}, S. 2022, \bibinfo{title}{{Binary parameters from astrometric and spectroscopic errors - candidate hierarchical triples and massive dark companions in Gaia DR3},} \mnras, 516, 3661, \dodoi{10.1093/mnras/stac2532}

% type= article
\bibitem[{J.~J. {Andrews} {et~al.}(2022){Andrews}, {Taggart}, \& {Foley}}]{Andrews:2022}
{Andrews}, J.~J., {Taggart}, K., \& {Foley}, R. 2022, \bibinfo{title}{{A Sample of Neutron Star and Black Hole Binaries Detected through Gaia DR3 Astrometry},} arXiv e-prints, arXiv:2207.00680, \dodoi{10.48550/arXiv.2207.00680}

% type= article
\bibitem[{E. {Balbinot} {et~al.}(2024){Balbinot}, {Dodd}, {Matsuno}, {Lardo}, {Helmi}, {Panuzzo}, {Mazeh}, {Holl}, {Caffau}, {Jorissen}, {Babusiaux}, {Gavras}, {Wyrzykowski}, {Eyer}, {Leclerc}, {Bombrun}, {Mowlavi}, {Seabroke}, {Cabrera-Ziri}, {Callingham}, {Ruiz-Lara}, \& {Starkenburg}}]{Balbinot:2024}
{Balbinot}, E., {Dodd}, E., {Matsuno}, T., {et~al.} 2024, \bibinfo{title}{{The 33 M$_{{\ensuremath{\odot}}}$ black hole Gaia BH3 is part of the disrupted ED-2 star cluster},} \aap, 687, L3, \dodoi{10.1051/0004-6361/202450425}

% type= article
\bibitem[{T.~G. {Barnes} {et~al.}(2021){Barnes}, {Guggenberger}, \& {Kolenberg}}]{Barnes:2021}
{Barnes}, III, T.~G., {Guggenberger}, E., \& {Kolenberg}, K. 2021, \bibinfo{title}{{A Radial-velocity Search for Binary RR Lyrae Variables},} \aj, 162, 117, \dodoi{10.3847/1538-3881/ac09f2}

% type= article
\bibitem[{K. {Belczynski} {et~al.}(2010){Belczynski}, {Dominik}, {Bulik}, {O'Shaughnessy}, {Fryer}, \& {Holz}}]{Belczynski:2010}
{Belczynski}, K., {Dominik}, M., {Bulik}, T., {et~al.} 2010, \bibinfo{title}{{The Effect of Metallicity on the Detection Prospects for Gravitational Waves},} \apjl, 715, L138, \dodoi{10.1088/2041-8205/715/2/L138}

% type= article
\bibitem[{V. {Belokurov} {et~al.}(2020){Belokurov}, {Penoyre}, {Oh}, {Iorio}, {Hodgkin}, {Evans}, {Everall}, {Koposov}, {Tout}, {Izzard}, {Clarke}, \& {Brown}}]{Belokurov:2020}
{Belokurov}, V., {Penoyre}, Z., {Oh}, S., {et~al.} 2020, \bibinfo{title}{{Unresolved stellar companions with Gaia DR2 astrometry},} \mnras, 496, 1922, \dodoi{10.1093/mnras/staa1522}

% type= inproceedings
\bibitem[{R. {Bernstein} {et~al.}(2003){Bernstein}, {Shectman}, {Gunnels}, {Mochnacki}, \& {Athey}}]{Bernstein:2003}
{Bernstein}, R., {Shectman}, S.~A., {Gunnels}, S.~M., {Mochnacki}, S., \& {Athey}, A.~E. 2003, \bibinfo{title}{{MIKE: A Double Echelle Spectrograph for the Magellan Telescopes at Las Campanas Observatory},} in Society of Photo-Optical Instrumentation Engineers (SPIE) Conference Series, Vol. 4841, Instrument Design and Performance for Optical/Infrared Ground-based Telescopes, ed. M.~{Iye} \& A.~F.~M. {Moorwood}, 1694--1704, \dodoi{10.1117/12.461502}

% type= article
\bibitem[{R. {Blomme} {et~al.}(2023){Blomme}, {Fr{\'e}mat}, {Sartoretti}, {Guerrier}, {Panuzzo}, {Katz}, {Seabroke}, {Th{\'e}venin}, {Cropper}, {Benson}, {Damerdji}, {Haigron}, {Marchal}, {Smith}, {Baker}, {Chemin}, {David}, {Dolding}, {Gosset}, {Jan{\ss}en}, {Jasniewicz}, {Lobel}, {Plum}, {Samaras}, {Snaith}, {Soubiran}, {Vanel}, {Zwitter}, {Brouillet}, {Caffau}, {Crifo}, {Fabre}, {Fragkoudi}, {Huckle}, {Jean-Antoine Piccolo}, {Lasne}, {Leclerc}, {Mastrobuono-Battisti}, {Royer}, {Viala}, \& {Zorec}}]{Blomme:2023}
{Blomme}, R., {Fr{\'e}mat}, Y., {Sartoretti}, P., {et~al.} 2023, \bibinfo{title}{{Gaia Data Release 3. Hot-star radial velocities},} \aap, 674, A7, \dodoi{10.1051/0004-6361/202243685}

% type= article
\bibitem[{J. {Bodensteiner} {et~al.}(2020){Bodensteiner}, {Shenar}, {Mahy}, {Fabry}, {Marchant}, {Abdul-Masih}, {Banyard}, {Bowman}, {Dsilva}, {Frost}, {Hawcroft}, {Reggiani}, \& {Sana}}]{Bodensteiner:2020}
{Bodensteiner}, J., {Shenar}, T., {Mahy}, L., {et~al.} 2020, \bibinfo{title}{{Is HR 6819 a triple system containing a black hole?. An alternative explanation},} \aap, 641, A43, \dodoi{10.1051/0004-6361/202038682}

% type= article
\bibitem[{J. {Bovy} {et~al.}(2019){Bovy}, {Leung}, {Hunt}, {Mackereth}, {Garcia-Hernandez}, \& {Roman-Lopes}}]{Bovy:2019}
{Bovy}, J., {Leung}, H.~W., {Hunt}, J. A.~S., {et~al.} 2019, \bibinfo{title}{{Life in the fast lane: a direct view of the dynamics, formation, and evolution of the Milky Way's bar},} arXiv e-prints, arXiv:1905.11404.
\newblock \doarXiv{1905.11404}

% type= article
\bibitem[{S. {Chakrabarti} {et~al.}(2023){Chakrabarti}, {Simon}, {Craig}, {Reggiani}, {Brandt}, {Guhathakurta}, {Dalba}, {Kirby}, {Chang}, {Hey}, {Savino}, {Geha}, \& {Thompson}}]{Chakrabarti:2023}
{Chakrabarti}, S., {Simon}, J.~D., {Craig}, P.~A., {et~al.} 2023, \bibinfo{title}{{A Noninteracting Galactic Black Hole Candidate in a Binary System with a Main-sequence Star},} \aj, 166, 6, \dodoi{10.3847/1538-3881/accf21}

% type= article
\bibitem[{M. {Cropper} {et~al.}(2018){Cropper}, {Katz}, {Sartoretti}, {Prusti}, {de Bruijne}, {Chassat}, {Charvet}, {Boyadjian}, {Perryman}, {Sarri}, {Gare}, {Erdmann}, {Munari}, {Zwitter}, {Wilkinson}, {Arenou}, {Vallenari}, {G{\'o}mez}, {Panuzzo}, {Seabroke}, {Allende Prieto}, {Benson}, {Marchal}, {Huckle}, {Smith}, {Dolding}, {Jan{\ss}en}, {Viala}, {Blomme}, {Baker}, {Boudreault}, {Crifo}, {Soubiran}, {Fr{\'e}mat}, {Jasniewicz}, {Guerrier}, {Guy}, {Turon}, {Jean-Antoine-Piccolo}, {Th{\'e}venin}, {David}, {Gosset}, \& {Damerdji}}]{Cropper:2018}
{Cropper}, M., {Katz}, D., {Sartoretti}, P., {et~al.} 2018, \bibinfo{title}{{Gaia Data Release 2. Gaia Radial Velocity Spectrometer},} \aap, 616, A5, \dodoi{10.1051/0004-6361/201832763}

% type= article
\bibitem[{X.-Q. {Cui} {et~al.}(2012){Cui}, {Zhao}, {Chu}, {Li}, {Li}, {Zhang}, {Su}, {Yao}, {Wang}, {Xing}, {Li}, {Zhu}, {Wang}, {Gu}, {Luo}, {Xu}, {Zhang}, {Liu}, {Zhang}, {Yang}, {Cao}, {Chen}, {Chen}, {Chen}, {Chen}, {Chu}, {Feng}, {Gong}, {Hou}, {Hu}, {Hu}, {Hu}, {Jia}, {Jiang}, {Jiang}, {Jiang}, {Jin}, {Li}, {Li}, {Li}, {Liu}, {Liu}, {Lu}, {Mao}, {Men}, {Qi}, {Qi}, {Shi}, {Tang}, {Tao}, {Wang}, {Wang}, {Wang}, {Wang}, {Wang}, {Wang}, {Wang}, {Wang}, {Wang}, {Wang}, {Wang}, {Wang}, {Xu}, {Xu}, {Yang}, {Yu}, {Yuan}, {Yuan}, {Zhai}, {Zhang}, {Zhang}, {Zhang}, {Zhao}, {Zhou}, {Zhou}, {Zhu}, \& {Zou}}]{Cui:2012}
{Cui}, X.-Q., {Zhao}, Y.-H., {Chu}, Y.-Q., {et~al.} 2012, \bibinfo{title}{{The Large Sky Area Multi-Object Fiber Spectroscopic Telescope (LAMOST)},} Research in Astronomy and Astrophysics, 12, 1197, \dodoi{10.1088/1674-4527/12/9/003}

% type= article
\bibitem[{R. {Drimmel} {et~al.}(2003){Drimmel}, {Cabrera-Lavers}, \& {L{\'o}pez-Corredoira}}]{Drimmel:2003}
{Drimmel}, R., {Cabrera-Lavers}, A., \& {L{\'o}pez-Corredoira}, M. 2003, \bibinfo{title}{{A three-dimensional Galactic extinction model},} \aap, 409, 205, \dodoi{10.1051/0004-6361:20031070}

% type= article
\bibitem[{K. {El-Badry}(2024{\natexlab{a}}){El-Badry}}]{El-Badry:2024a}
{El-Badry}, K. 2024{\natexlab{a}}, \bibinfo{title}{{On the formation of a 33 solar-mass black hole in a low-metallicity binary},} The Open Journal of Astrophysics, 7, 38, \dodoi{10.33232/001c.117652}

% type= article
\bibitem[{K. {El-Badry}(2024{\natexlab{b}}){El-Badry}}]{El-Badry:2024}
{El-Badry}, K. 2024{\natexlab{b}}, \bibinfo{title}{{On the formation of a 33 solar-mass black hole in a low-metallicity binary},} The Open Journal of Astrophysics, 7, 38, \dodoi{10.33232/001c.117652}

% type= article
\bibitem[{K. {El-Badry} \& K.~B. {Burdge}(2022){El-Badry} \& {Burdge}}]{El-Badry:2022a}
{El-Badry}, K., \& {Burdge}, K.~B. 2022, \bibinfo{title}{{NGC 1850 BH1 is another stripped-star binary masquerading as a black hole},} \mnras, 511, 24, \dodoi{10.1093/mnrasl/slab135}

% type= article
\bibitem[{K. {El-Badry} {et~al.}(2024{\natexlab{a}}){El-Badry}, {Lam}, {Holl}, {Halbwachs}, {Rix}, {Mazeh}, \& {Shahaf}}]{El-Badry:2024c}
{El-Badry}, K., {Lam}, C., {Holl}, B., {et~al.} 2024{\natexlab{a}}, \bibinfo{title}{{A generative model for Gaia astrometric orbit catalogs: selection functions for binary stars, giant planets, and compact object companions},} The Open Journal of Astrophysics, 7, 100, \dodoi{10.33232/001c.125461}

% type= article
\bibitem[{K. {El-Badry} \& H.-W. {Rix}(2022){El-Badry} \& {Rix}}]{El-Badry:2022}
{El-Badry}, K., \& {Rix}, H.-W. 2022, \bibinfo{title}{{What are the spectroscopic binaries with high-mass functions near the Gaia DR3 main sequence?},} \mnras, 515, 1266, \dodoi{10.1093/mnras/stac1797}

% type= article
\bibitem[{K. {El-Badry} {et~al.}(2022){El-Badry}, {Seeburger}, {Jayasinghe}, {Rix}, {Almada}, {Conroy}, {Price-Whelan}, \& {Burdge}}]{El-Badry:2022b}
{El-Badry}, K., {Seeburger}, R., {Jayasinghe}, T., {et~al.} 2022, \bibinfo{title}{{Unicorns and giraffes in the binary zoo: stripped giants with subgiant companions},} \mnras, 512, 5620, \dodoi{10.1093/mnras/stac815}

% type= article
\bibitem[{K. {El-Badry} {et~al.}(2023{\natexlab{a}}){El-Badry}, {Rix}, {Quataert}, {Howard}, {Isaacson}, {Fuller}, {Hawkins}, {Breivik}, {Wong}, {Rodriguez}, {Conroy}, {Shahaf}, {Mazeh}, {Arenou}, {Burdge}, {Bashi}, {Faigler}, {Weisz}, {Seeburger}, {Almada Monter}, \& {Wojno}}]{El-Badry:2023a}
{El-Badry}, K., {Rix}, H.-W., {Quataert}, E., {et~al.} 2023{\natexlab{a}}, \bibinfo{title}{{A Sun-like star orbiting a black hole},} \mnras, 518, 1057, \dodoi{10.1093/mnras/stac3140}

% type= article
\bibitem[{K. {El-Badry} {et~al.}(2023{\natexlab{b}}){El-Badry}, {Rix}, {Cendes}, {Rodriguez}, {Conroy}, {Quataert}, {Hawkins}, {Zari}, {Hobson}, {Breivik}, {Rau}, {Berger}, {Shahaf}, {Seeburger}, {Burdge}, {Latham}, {Buchhave}, {Bieryla}, {Bashi}, {Mazeh}, \& {Faigler}}]{El-Badry:2023b}
{El-Badry}, K., {Rix}, H.-W., {Cendes}, Y., {et~al.} 2023{\natexlab{b}}, \bibinfo{title}{{A red giant orbiting a black hole},} \mnras, 521, 4323, \dodoi{10.1093/mnras/stad799}

% type= article
\bibitem[{K. {El-Badry} {et~al.}(2024{\natexlab{b}}){El-Badry}, {Rix}, {Latham}, {Shahaf}, {Mazeh}, {Bieryla}, {Buchhave}, {Andrae}, {Yamaguchi}, {Isaacson}, {Howard}, {Savino}, \& {Ilyin}}]{El-Badry:2024b}
{El-Badry}, K., {Rix}, H.-W., {Latham}, D.~W., {et~al.} 2024{\natexlab{b}}, \bibinfo{title}{{A population of neutron star candidates in wide orbits from Gaia astrometry},} The Open Journal of Astrophysics, 7, 58, \dodoi{10.33232/001c.121261}

% type= article
\bibitem[{K. {El-Badry} {et~al.}(2024{\natexlab{c}}){El-Badry}, {Simon}, {Reggiani}, {Rix}, {Latham}, {Bieryla}, {Buchhave}, {Shahaf}, {Mazeh}, {Chakrabarti}, {Guhathakurta}, {Ilyin}, \& {Tauris}}]{El-Badry:2024d}
{El-Badry}, K., {Simon}, J.~D., {Reggiani}, H., {et~al.} 2024{\natexlab{c}}, \bibinfo{title}{{A 1.9 solar-mass neutron star candidate in a 2-year orbit},} The Open Journal of Astrophysics, 7, 27, \dodoi{10.33232/001c.116675}

% type= article
\bibitem[{W.~M. {Farr} {et~al.}(2011){Farr}, {Sravan}, {Cantrell}, {Kreidberg}, {Bailyn}, {Mandel}, \& {Kalogera}}]{Farr:2011}
{Farr}, W.~M., {Sravan}, N., {Cantrell}, A., {et~al.} 2011, \bibinfo{title}{{The Mass Distribution of Stellar-mass Black Holes},} \apj, 741, 103, \dodoi{10.1088/0004-637X/741/2/103}

% type= article
\bibitem[{ {Gaia Collaboration} {et~al.}(2023{\natexlab{a}}){Gaia Collaboration}, {Arenou}, {Babusiaux}, {Barstow}, {Faigler}, {Jorissen}, {Kervella}, {Mazeh}, {Mowlavi}, {Panuzzo}, {Sahlmann}, {Shahaf}, {Sozzetti}, {Bauchet}, {Damerdji}, {Gavras}, {Giacobbe}, {Gosset}, {Halbwachs}, {Holl}, {Lattanzi}, {Leclerc}, {Morel}, {Pourbaix}, {Re Fiorentin}, {Sadowski}, {S{\'e}gransan}, {Siopis}, {Teyssier}, {Zwitter}, {Planquart}, \& {Brown}}]{Arenou:2023}
{Gaia Collaboration}, {Arenou}, F., {Babusiaux}, C., {et~al.} 2023{\natexlab{a}}, \bibinfo{title}{{Gaia Data Release 3. Stellar multiplicity, a teaser for the hidden treasure},} \aap, 674, A34, \dodoi{10.1051/0004-6361/202243782}

% type= article
\bibitem[{ {Gaia Collaboration} {et~al.}(2023{\natexlab{b}}){Gaia Collaboration}, {Trabucchi}, {Mowlavi}, {Lebzelter}, {Lecoeur-Taibi}, {Audard}, {Eyer}, {Garc{\'\i}a-Lario}, {Gavras}, {Holl}, {Jevardat de Fombelle}, {Nienartowicz}, {Rimoldini}, {Sartoretti}, {Blomme}, {Fr{\'e}mat}, {Marchal}, {Damerdji}, {Brown}, {Guerrier}, {Panuzzo}, {Katz}, {Seabroke}, {Benson}, {Haigron}, {Smith}, {Lobel}, {Vallenari}, {Prusti}, {de Bruijne}, {Arenou}, {Babusiaux}, {Barbier}, {Biermann}, {Creevey}, {Ducourant}, {Evans}, {Guerra}, {Hutton}, {Jordi}, {Klioner}, {Lammers}, {Lindegren}, {Luri}, {Mignard}, {Randich}, {Smiljanic}, {Tanga}, {Walton}, {Bailer-Jones}, {Bastian}, {Cropper}, {Drimmel}, {Lattanzi}, {Soubiran}, {van Leeuwen}, {Bakker}, {Casta{\~n}eda}, {De Angeli}, {Fabricius}, {Fouesneau}, {Galluccio}, {Masana}, {Messineo}, {Nicolas}, {Pailler}, {Riclet}, {Roux}, {Sordo}, {Th{\'e}venin}, {Gracia-Abril}, {Portell}, {Teyssier}, {Altmann}, {Berthier}, {Burgess}, {Busonero}, {Busso}, {C{\'a}novas}, {Carry}, {Cheek},
  {Clementini}, {Davidson}, {de Teodoro}, {Delchambre}, {Dell'Oro}, {Fraile Garcia}, {Garabato}, {Garralda Torres}, {Hambly}, {Harrison}, {Hatzidimitriou}, {Hern{\'a}ndez}, {Hodgkin}, {Jamal}, {Jordan}, {Krone-Martins}, {Lanzafame}, {L{\"o}ffler}, {Lorca}, {Marrese}, {Moitinho}, {Muinonen}, {Nu{\~n}ez Campos}, {Oreshina-Slezak}, {Osborne}, {Pancino}, {Pauwels}, {Recio-Blanco}, {Riello}, {Robin}, {Roegiers}, {Sarro}, {Schultheis}, {Siopis}, {Sozzetti}, {Utrilla}, {van Leeuwen}, {Weingrill}, {Abbas}, {{\'A}brah{\'a}m}, {Abreu Aramburu}, {Aerts}, {Altavilla}, {{\'A}lvarez}, {Alves}, {Anders}, {Anderson}, {Antoja}, {Baines}, {Baker}, {Balog}, {Barache}, {Barbato}, {Barros}, {Barstow}, {Bartolom{\'e}}, {Bashi}, {Bauchet}, {Baudeau}, {Becciani}, {Bedin}, {Bellas-Velidis}, {Bellazzini}, {Beordo}, {Berihuete}, {Bernet}, {Bertolotto}, {Bertone}, {Bianchi}, {Binnenfeld}, {Blazere}, {Boch}, {Bombrun}, {Bouquillon}, {Bragaglia}, {Braine}, {Bramante}, {Breedt}, {Bressan}, {Brouillet}, {Brugaletta}, {Bucciarelli},
  {Butkevich}, {Buzzi}, {Caffau}, {Cancelliere}, {Cannizzo}, {Carballo}, {Carlucci}, {Carnerero}, {Carrasco}, {Carretero}, {Carton}, {Casamiquela}, {Castellani}, {Castro-Ginard}, {Cesare}, {Charlot}, {Chemin}, {Chiaramida}, {Chiavassa}, {Chornay}, {Collins}, {Contursi}, {Cooper}, {Cornez}, {Crosta}, {Crowley}, {Dafonte}, {David}, {de Laverny}, {De Luise}, {De March}, {De Ridder}, {de Souza}, {de Torres}, {del Peloso}, {Delbo}, \& {Delgado}}]{Trabucchi:2023}
{Gaia Collaboration}, {Trabucchi}, M., {Mowlavi}, N., {et~al.} 2023{\natexlab{b}}, \bibinfo{title}{{Gaia Focused Product Release: Radial velocity time series of long-period variables},} \aap, 680, A36, \dodoi{10.1051/0004-6361/202347287}

% type= article
\bibitem[{ {Gaia Collaboration} {et~al.}(2024){Gaia Collaboration}, {Panuzzo}, {Mazeh}, {Arenou}, {Holl}, {Caffau}, {Jorissen}, {Babusiaux}, {Gavras}, {Sahlmann}, {Bastian}, {Wyrzykowski}, {Eyer}, {Leclerc}, {Bauchet}, {Bombrun}, {Mowlavi}, {Seabroke}, {Teyssier}, {Balbinot}, {Helmi}, {Brown}, {Vallenari}, {Prusti}, {de Bruijne}, {Barbier}, {Biermann}, {Creevey}, {Ducourant}, {Evans}, {Guerra}, {Hutton}, {Jordi}, {Klioner}, {Lammers}, {Lindegren}, {Luri}, {Mignard}, {Nicolas}, {Randich}, {Sartoretti}, {Smiljanic}, {Tanga}, {Walton}, {Aerts}, {Bailer-Jones}, {Cropper}, {Drimmel}, {Jansen}, {Katz}, {Lattanzi}, {Soubiran}, {Th{\'e}venin}, {van Leeuwen}, {Andrae}, {Audard}, {Bakker}, {Blomme}, {Casta{\~n}eda}, {De Angeli}, {Fabricius}, {Fouesneau}, {Fr{\'e}mat}, {Galluccio}, {Guerrier}, {Heiter}, {Masana}, {Messineo}, {Nienartowicz}, {Pailler}, {Riclet}, {Roux}, {Sordo}, {Gracia-Abril}, {Portell}, {Altmann}, {Benson}, {Berthier}, {Burgess}, {Busonero}, {Busso}, {Cacciari}, {C{\'a}novas}, {Carrasco}, {Carry},
  {Cellino}, {Cheek}, {Clementini}, {Damerdji}, {Davidson}, {de Teodoro}, {Delchambre}, {Dell'Oro}, {Fraile Garcia}, {Garabato}, {Garc{\'\i}a-Lario}, {Haigron}, {Hambly}, {Harrison}, {Hatzidimitriou}, {Hern{\'a}ndez}, {Hestroffer}, {Hodgkin}, {Jamal}, {Jevardat de Fombelle}, {Jordan}, {Krone-Martins}, {Lanzafame}, {L{\"o}ffler}, {Lorca}, {Marchal}, {Marrese}, {Moitinho}, {Muinonen}, {Nu{\~n}ez Campos}, {Oreshina-Slezak}, {Osborne}, {Pancino}, {Pauwels}, {Recio-Blanco}, {Riello}, {Rimoldini}, {Robin}, {Roegiers}, {Sarro}, \& {Schultheis}}]{Panuzzo:2024}
{Gaia Collaboration}, {Panuzzo}, P., {Mazeh}, T., {et~al.} 2024, \bibinfo{title}{{Discovery of a dormant 33 solar-mass black hole in pre-release Gaia astrometry},} \aap, 686, L2, \dodoi{10.1051/0004-6361/202449763}

% type= article
\bibitem[{S. {Geier} {et~al.}(2023){Geier}, {Dorsch}, {Dawson}, {Pelisoli}, {Munday}, {Marsh}, {Schaffenroth}, \& {Heber}}]{Geier:2023}
{Geier}, S., {Dorsch}, M., {Dawson}, H., {et~al.} 2023, \bibinfo{title}{{The first massive compact companion in a wide orbit around a hot subdwarf star},} \aap, 677, A11, \dodoi{10.1051/0004-6361/202346407}

% type= article
\bibitem[{C. {Gielen} {et~al.}(2011){Gielen}, {Bouwman}, {van Winckel}, {Lloyd Evans}, {Woods}, {Kemper}, {Marengo}, {Meixner}, {Sloan}, \& {Tielens}}]{Gielen:2011}
{Gielen}, C., {Bouwman}, J., {van Winckel}, H., {et~al.} 2011, \bibinfo{title}{{Silicate features in Galactic and extragalactic post-AGB discs},} \aap, 533, A99, \dodoi{10.1051/0004-6361/201117364}

% type= article
\bibitem[{E. {Gosset} {et~al.}(2025){Gosset}, {Damerdji}, {Morel}, {Delchambre}, {Halbwachs}, {Sadowski}, {Pourbaix}, {Sozzetti}, {Panuzzo}, \& {Arenou}}]{Gosset:2025}
{Gosset}, E., {Damerdji}, Y., {Morel}, T., {et~al.} 2025, \bibinfo{title}{{Gaia Data Release 3: Spectroscopic binary-star orbital solutions: The SB1 processing chain},} \aap, 693, A124, \dodoi{10.1051/0004-6361/202450600}

% type= article
\bibitem[{G.~M. {Green} {et~al.}(2019){Green}, {Schlafly}, {Zucker}, {Speagle}, \& {Finkbeiner}}]{Green:2019}
{Green}, G.~M., {Schlafly}, E., {Zucker}, C., {Speagle}, J.~S., \& {Finkbeiner}, D. 2019, \bibinfo{title}{{A 3D Dust Map Based on Gaia, Pan-STARRS 1, and 2MASS},} \apj, 887, 93, \dodoi{10.3847/1538-4357/ab5362}

% type= article
\bibitem[{J.-L. {Halbwachs} {et~al.}(2023){Halbwachs}, {Pourbaix}, {Arenou}, {Galluccio}, {Guillout}, {Bauchet}, {Marchal}, {Sadowski}, \& {Teyssier}}]{Halbwachs:2023}
{Halbwachs}, J.-L., {Pourbaix}, D., {Arenou}, F., {et~al.} 2023, \bibinfo{title}{{Gaia Data Release 3. Astrometric binary star processing},} \aap, 674, A9, \dodoi{10.1051/0004-6361/202243969}

% type= misc
\bibitem[{N. {Hambly} {et~al.}(2022){Hambly}, {Andrae}, {De Angeli}, {Antonio}, {Arenou}, {Audard}, {Babusiaux}, {Bailer-Jones}, {Bakker}, {Bastian}, {Bauchet}, {Bellas-Velidis}, {Blomme}, {Bombrun}, {Brouillet}, {Brugaletta}, {de Bruijne}, {Busonero}, {Busso}, {Carballo}, {Carnerero}, {Clementini}, {Creevey}, {Damerdji}, {Delchambre}, {Distefano}, {Drimmel}, {Ducourant}, {Duran}, {Fabricius}, {Eyer}, {Faigler}, {Findeisen}, {Jevardat de Fombelle}, {Fouesneau}, {Fr{\'e}mat}, {Galluccio}, {Garabato}, {Gavras}, {Giuffrida}, {Gomel}, {Gonz{\'a}lez}, {Gonz{\'a}lez-N{\'u}{\~n}ez}, {Gosset}, {Gracia-Abril}, {Halbwachs}, {Harrison}, {Heiter}, {Hernandez}, {Hestroffer}, {Hobbs}, {Hodgkin}, {Holl}, {Hutton}, {Katz}, {Klioner}, {Leccia}, {Lebreton}, {Lecoeur-Ta{\"\i}bi}, {van Leeuwen}, {Lindegren}, {Lobel}, {Luri}, {Mantelet}, {Marrese}, {Marinoni}, {Marshall}, {Masana}, {Mazeh}, {Michalik}, {Molinaro}, {Mora}, {Mowlavi}, {Nienartowicz}, {Ordenovic}, {Panahi}, {Pancino}, {Pauwels}, {Pichon}, {Portell}, {Pourbaix},
  {Raiteri}, {Recio-Blanco}, {De Ridder}, {Riello}, {Rimoldini}, {Ripepi}, {Rixon}, {Robin}, {Rybizki}, {Sartoretti}, {Sarro Baro}, {Seabroke}, {Segovia Serrato}, {Siopis}, {Smart}, {Soubiran}, {Sozzetti}, {Spoto}, {Tanga}, {Teyssier}, {Utrilla}, {Masip Vela}, {Wyrzykowski}, \& {Zucker}}]{Hambly:2022}
{Hambly}, N., {Andrae}, R., {De Angeli}, F., {et~al.} 2022, {Gaia DR3 documentation Chapter 20: Datamodel description},

% type= article
\bibitem[{A. {Heger} {et~al.}(2003){Heger}, {Fryer}, {Woosley}, {Langer}, \& {Hartmann}}]{Heger:2003}
{Heger}, A., {Fryer}, C.~L., {Woosley}, S.~E., {Langer}, N., \& {Hartmann}, D.~H. 2003, \bibinfo{title}{{How Massive Single Stars End Their Life},} \apj, 591, 288, \dodoi{10.1086/375341}

% type= article
\bibitem[{A.~W. {Howard} {et~al.}(2010){Howard}, {Johnson}, {Marcy}, {Fischer}, {Wright}, {Bernat}, {Henry}, {Peek}, {Isaacson}, {Apps}, {Endl}, {Cochran}, {Valenti}, {Anderson}, \& {Piskunov}}]{Howard:2010}
{Howard}, A.~W., {Johnson}, J.~A., {Marcy}, G.~W., {et~al.} 2010, \bibinfo{title}{{The California Planet Survey. I. Four New Giant Exoplanets},} \apj, 721, 1467, \dodoi{10.1088/0004-637X/721/2/1467}

% type= article
\bibitem[{S. {Janssens} {et~al.}(2022){Janssens}, {Shenar}, {Sana}, {Faigler}, {Langer}, {Marchant}, {Mazeh}, {Sch{\"u}rmann}, \& {Shahaf}}]{Janssens:2022}
{Janssens}, S., {Shenar}, T., {Sana}, H., {et~al.} 2022, \bibinfo{title}{{Uncovering astrometric black hole binaries with massive main-sequence companions with Gaia},} \aap, 658, A129, \dodoi{10.1051/0004-6361/202141866}

% type= article
\bibitem[{T. {Jayasinghe} {et~al.}(2023){Jayasinghe}, {Rowan}, {Thompson}, {Kochanek}, \& {Stanek}}]{Jayasinghe:2023}
{Jayasinghe}, T., {Rowan}, D.~M., {Thompson}, T.~A., {Kochanek}, C.~S., \& {Stanek}, K.~Z. 2023, \bibinfo{title}{{A search for compact object companions to high mass function single-lined spectroscopic binaries in Gaia DR3},} \mnras, 521, 5927, \dodoi{10.1093/mnras/stad909}

% type= article
\bibitem[{D. {Katz} {et~al.}(2023){Katz}, {Sartoretti}, {Guerrier}, {Panuzzo}, {Seabroke}, {Th{\'e}venin}, {Cropper}, {Benson}, {Blomme}, {Haigron}, {Marchal}, {Smith}, {Baker}, {Chemin}, {Damerdji}, {David}, {Dolding}, {Fr{\'e}mat}, {Gosset}, {Jan{\ss}en}, {Jasniewicz}, {Lobel}, {Plum}, {Samaras}, {Snaith}, {Soubiran}, {Vanel}, {Zwitter}, {Antoja}, {Arenou}, {Babusiaux}, {Brouillet}, {Caffau}, {Di Matteo}, {Fabre}, {Fabricius}, {Fragkoudi}, {Haywood}, {Huckle}, {Hottier}, {Lasne}, {Leclerc}, {Mastrobuono-Battisti}, {Royer}, {Teyssier}, {Zorec}, {Crifo}, {Jean-Antoine Piccolo}, {Turon}, \& {Viala}}]{Katz:2023}
{Katz}, D., {Sartoretti}, P., {Guerrier}, A., {et~al.} 2023, \bibinfo{title}{{Gaia Data Release 3. Properties and validation of the radial velocities},} \aap, 674, A5, \dodoi{10.1051/0004-6361/202244220}

% type= article
\bibitem[{D.~D. {Kelson}(2003){Kelson}}]{Kelson:2003}
{Kelson}, D.~D. 2003, \bibinfo{title}{{Optimal Techniques in Two-dimensional Spectroscopy: Background Subtraction for the 21st Century},} \pasp, 115, 688, \dodoi{10.1086/375502}

% type= article
\bibitem[{D.~D. {Kelson} {et~al.}(2000){Kelson}, {Illingworth}, {van Dokkum}, \& {Franx}}]{Kelson:2000}
{Kelson}, D.~D., {Illingworth}, G.~D., {van Dokkum}, P.~G., \& {Franx}, M. 2000, \bibinfo{title}{{The Evolution of Early-Type Galaxies in Distant Clusters. II. Internal Kinematics of 55 Galaxies in the z=0.33 Cluster CL 1358+62},} \apj, 531, 159, \dodoi{10.1086/308445}

% type= article
\bibitem[{C.~Y. {Lam} {et~al.}(2025){Lam}, {El-Badry}, \& {Simon}}]{Lam:2025}
{Lam}, C.~Y., {El-Badry}, K., \& {Simon}, J.~D. 2025, \bibinfo{title}{{A Fast, Analytic Empirical Model of the Gaia Data Release 3 Astrometric Orbit Catalog Selection Function},} \apj, 987, 215, \dodoi{10.3847/1538-4357/addac2}

% type= article
\bibitem[{C.~Y. {Lam} \& J.~R. {Lu}(2023){Lam} \& {Lu}}]{Lam:2023}
{Lam}, C.~Y., \& {Lu}, J.~R. 2023, \bibinfo{title}{{A Reanalysis of the Isolated Black Hole Candidate OGLE-2011-BLG-0462/MOA-2011-BLG-191},} \apj, 955, 116, \dodoi{10.3847/1538-4357/aced4a}

% type= article
\bibitem[{T. {Lebzelter} {et~al.}(2023){Lebzelter}, {Mowlavi}, {Lecoeur-Taibi}, {Trabucchi}, {Audard}, {Garc{\'\i}a-Lario}, {Gavras}, {Holl}, {Jevardat de Fombelle}, {Nienartowicz}, {Rimoldini}, \& {Eyer}}]{Lebzelter:2023}
{Lebzelter}, T., {Mowlavi}, N., {Lecoeur-Taibi}, I., {et~al.} 2023, \bibinfo{title}{{Gaia Data Release 3. The second Gaia catalogue of long-period variable candidates},} \aap, 674, A15, \dodoi{10.1051/0004-6361/202244241}

% type= unpublished
\bibitem[{L. Lindegren(2018)Lindegren}]{Lindegren:2018}
Lindegren, L. 2018, \bibinfo{title}{{R}e-normalising the astrometric chi-square in {G}aia {D}{R}2,} \url{http://www.rssd.esa.int/doc_fetch.php?id=3757412}

% type= article
\bibitem[{T. {Maas} {et~al.}(2002){Maas}, {Van Winckel}, \& {Waelkens}}]{Maas:2002}
{Maas}, T., {Van Winckel}, H., \& {Waelkens}, C. 2002, \bibinfo{title}{{RU Cen and SX Cen: Two strongly depleted RV Tauri stars in binary systems. The RV Tauri photometric b phenomenon and binarity},} \aap, 386, 504, \dodoi{10.1051/0004-6361:20020209}

% type= article
\bibitem[{S.~R. {Majewski} {et~al.}(2017){Majewski}, {Schiavon}, {Frinchaboy}, {Allende Prieto}, {Barkhouser}, {Bizyaev}, {Blank}, {Brunner}, {Burton}, {Carrera}, {Chojnowski}, {Cunha}, {Epstein}, {Fitzgerald}, {Garc{\'\i}a P{\'e}rez}, {Hearty}, {Henderson}, {Holtzman}, {Johnson}, {Lam}, {Lawler}, {Maseman}, {M{\'e}sz{\'a}ros}, {Nelson}, {Nguyen}, {Nidever}, {Pinsonneault}, {Shetrone}, {Smee}, {Smith}, {Stolberg}, {Skrutskie}, {Walker}, {Wilson}, {Zasowski}, {Anders}, {Basu}, {Beland}, {Blanton}, {Bovy}, {Brownstein}, {Carlberg}, {Chaplin}, {Chiappini}, {Eisenstein}, {Elsworth}, {Feuillet}, {Fleming}, {Galbraith-Frew}, {Garc{\'\i}a}, {Garc{\'\i}a-Hern{\'a}ndez}, {Gillespie}, {Girardi}, {Gunn}, {Hasselquist}, {Hayden}, {Hekker}, {Ivans}, {Kinemuchi}, {Klaene}, {Mahadevan}, {Mathur}, {Mosser}, {Muna}, {Munn}, {Nichol}, {O'Connell}, {Parejko}, {Robin}, {Rocha-Pinto}, {Schultheis}, {Serenelli}, {Shane}, {Silva Aguirre}, {Sobeck}, {Thompson}, {Troup}, {Weinberg}, \& {Zamora}}]{Majewski:2017}
{Majewski}, S.~R., {Schiavon}, R.~P., {Frinchaboy}, P.~M., {et~al.} 2017, \bibinfo{title}{{The Apache Point Observatory Galactic Evolution Experiment (APOGEE)},} \aj, 154, 94, \dodoi{10.3847/1538-3881/aa784d}

% type= article
\bibitem[{D.~J. {Marshall} {et~al.}(2006){Marshall}, {Robin}, {Reyl{\'e}}, {Schultheis}, \& {Picaud}}]{Marshall:2006}
{Marshall}, D.~J., {Robin}, A.~C., {Reyl{\'e}}, C., {Schultheis}, M., \& {Picaud}, S. 2006, \bibinfo{title}{{Modelling the Galactic interstellar extinction distribution in three dimensions},} \aap, 453, 635, \dodoi{10.1051/0004-6361:20053842}

% type= article
\bibitem[{D.~C. {Martin} {et~al.}(2005){Martin}, {Fanson}, {Schiminovich}, {Morrissey}, {Friedman}, {Barlow}, {Conrow}, {Grange}, {Jelinsky}, {Milliard}, {Siegmund}, {Bianchi}, {Byun}, {Donas}, {Forster}, {Heckman}, {Lee}, {Madore}, {Malina}, {Neff}, {Rich}, {Small}, {Surber}, {Szalay}, {Welsh}, \& {Wyder}}]{Martin:2005}
{Martin}, D.~C., {Fanson}, J., {Schiminovich}, D., {et~al.} 2005, \bibinfo{title}{{The Galaxy Evolution Explorer: A Space Ultraviolet Survey Mission},} \apjl, 619, L1, \dodoi{10.1086/426387}

% type= article
\bibitem[{N. {Mowlavi} {et~al.}(2023){Mowlavi}, {Holl}, {Lecoeur-Ta{\"\i}bi}, {Barblan}, {Kochoska}, {Pr{\v{s}}a}, {Mazeh}, {Rimoldini}, {Gavras}, {Audard}, {Jevardat de Fombelle}, {Nienartowicz}, {Garc{\'\i}a-Lario}, \& {Eyer}}]{Mowlavi:2023}
{Mowlavi}, N., {Holl}, B., {Lecoeur-Ta{\"\i}bi}, I., {et~al.} 2023, \bibinfo{title}{{Gaia Data Release 3. The first Gaia catalogue of eclipsing-binary candidates},} \aap, 674, A16, \dodoi{10.1051/0004-6361/202245330}

% type= article
\bibitem[{J. {M{\"u}ller-Horn} {et~al.}(2025){M{\"u}ller-Horn}, {Rix}, {El-Badry}, {Pennell}, {Green}, {Li}, \& {Seeburger}}]{Muller-Horn:2025}
{M{\"u}ller-Horn}, J., {Rix}, H.-W., {El-Badry}, K., {et~al.} 2025, \bibinfo{title}{{Dormant BH candidates from Gaia DR3 summary diagnostics},} arXiv e-prints, arXiv:2510.05982, \dodoi{10.48550/arXiv.2510.05982}

% type= article
\bibitem[{P. {Nagarajan} {et~al.}(2025{\natexlab{a}}){Nagarajan}, {El-Badry}, {Reggiani}, {Lam}, {Simon}, {M{\"u}ller-Horn}, {Seeburger}, {Rix}, {Isaacson}, {Lu}, {Chandra}, \& {Andrae}}]{Nagarajan:2025b}
{Nagarajan}, P., {El-Badry}, K., {Reggiani}, H., {et~al.} 2025{\natexlab{a}}, \bibinfo{title}{{A Spectroscopic Search for Dormant Black Holes in Low-Metallicity Binaries},} arXiv e-prints, arXiv:2507.12532, \dodoi{10.48550/arXiv.2507.12532}

% type= article
\bibitem[{P. {Nagarajan} {et~al.}(2025{\natexlab{b}}){Nagarajan}, {El-Badry}, {Chawla}, {Di Carlo}, {Breivik}, {Rodriguez}, {Agrawal}, {Delfavero}, \& {Chatterjee}}]{Nagarajan:2025a}
{Nagarajan}, P., {El-Badry}, K., {Chawla}, C., {et~al.} 2025{\natexlab{b}}, \bibinfo{title}{{Realistic Predictions for Gaia Black Hole Discoveries: Comparison of Isolated Binary and Dynamical Formation Models},} \pasp, 137, 044202, \dodoi{10.1088/1538-3873/adc839}

% type= article
\bibitem[{G.-M. {Oomen} {et~al.}(2018){Oomen}, {Van Winckel}, {Pols}, {Nelemans}, {Escorza}, {Manick}, {Kamath}, \& {Waelkens}}]{Oomen:2018}
{Oomen}, G.-M., {Van Winckel}, H., {Pols}, O., {et~al.} 2018, \bibinfo{title}{{Orbital properties of binary post-AGB stars},} \aap, 620, A85, \dodoi{10.1051/0004-6361/201833816}

% type= article
\bibitem[{F. {{\"O}zel} {et~al.}(2010){{\"O}zel}, {Psaltis}, {Narayan}, \& {McClintock}}]{Ozel:2010}
{{\"O}zel}, F., {Psaltis}, D., {Narayan}, R., \& {McClintock}, J.~E. 2010, \bibinfo{title}{{The Black Hole Mass Distribution in the Galaxy},} \apj, 725, 1918, \dodoi{10.1088/0004-637X/725/2/1918}

% type= article
\bibitem[{Z. {Penoyre} {et~al.}(2020){Penoyre}, {Belokurov}, {Wyn Evans}, {Everall}, \& {Koposov}}]{Penoyre:2020}
{Penoyre}, Z., {Belokurov}, V., {Wyn Evans}, N., {Everall}, A., \& {Koposov}, S.~E. 2020, \bibinfo{title}{{Binary deviations from single object astrometry},} \mnras, 495, 321, \dodoi{10.1093/mnras/staa1148}

% type= article
\bibitem[{G.~R. {Ricker} {et~al.}(2015){Ricker}, {Winn}, {Vanderspek}, {Latham}, {Bakos}, {Bean}, {Berta-Thompson}, {Brown}, {Buchhave}, {Butler}, {Butler}, {Chaplin}, {Charbonneau}, {Christensen-Dalsgaard}, {Clampin}, {Deming}, {Doty}, {De Lee}, {Dressing}, {Dunham}, {Endl}, {Fressin}, {Ge}, {Henning}, {Holman}, {Howard}, {Ida}, {Jenkins}, {Jernigan}, {Johnson}, {Kaltenegger}, {Kawai}, {Kjeldsen}, {Laughlin}, {Levine}, {Lin}, {Lissauer}, {MacQueen}, {Marcy}, {McCullough}, {Morton}, {Narita}, {Paegert}, {Palle}, {Pepe}, {Pepper}, {Quirrenbach}, {Rinehart}, {Sasselov}, {Sato}, {Seager}, {Sozzetti}, {Stassun}, {Sullivan}, {Szentgyorgyi}, {Torres}, {Udry}, \& {Villasenor}}]{Ricker:2015}
{Ricker}, G.~R., {Winn}, J.~N., {Vanderspek}, R., {et~al.} 2015, \bibinfo{title}{{Transiting Exoplanet Survey Satellite (TESS)},} Journal of Astronomical Telescopes, Instruments, and Systems, 1, 014003, \dodoi{10.1117/1.JATIS.1.1.014003}

% type= article
\bibitem[{K.~C. {Sahu} {et~al.}(2025){Sahu}, {Anderson}, {Casertano}, {Bond}, {Dominik}, {Calamida}, {Bellini}, {Brown}, {Ferguson}, \& {Rejkuba}}]{Sahu:2025}
{Sahu}, K.~C., {Anderson}, J., {Casertano}, S., {et~al.} 2025, \bibinfo{title}{{OGLE-2011-BLG-0462: An Isolated Stellar-mass Black Hole Confirmed Using New HST Astrometry and Updated Photometry},} \apj, 983, 104, \dodoi{10.3847/1538-4357/adbe6e}

% type= article
\bibitem[{N.~N. {Samus'} {et~al.}(2017){Samus'}, {Kazarovets}, {Durlevich}, {Kireeva}, \& {Pastukhova}}]{Samus:2017}
{Samus'}, N.~N., {Kazarovets}, E.~V., {Durlevich}, O.~V., {Kireeva}, N.~N., \& {Pastukhova}, E.~N. 2017, \bibinfo{title}{{General catalogue of variable stars: Version GCVS 5.1},} Astronomy Reports, 61, 80, \dodoi{10.1134/S1063772917010085}

% type= article
\bibitem[{E. {Semkov}(2015){Semkov}}]{Semkov:2015}
{Semkov}, E. 2015, \bibinfo{title}{{The new FUor candidate V960 Mon (2MASS J06593158-0405277) still retains at high brightness level},} The Astronomer's Telegram, 8019, 1

% type= article
\bibitem[{S. {Shahaf} {et~al.}(2023){Shahaf}, {Bashi}, {Mazeh}, {Faigler}, {Arenou}, {El-Badry}, \& {Rix}}]{Shahaf:2023}
{Shahaf}, S., {Bashi}, D., {Mazeh}, T., {et~al.} 2023, \bibinfo{title}{{Triage of the Gaia DR3 astrometric orbits - I. A sample of binaries with probable compact companions},} \mnras, 518, 2991, \dodoi{10.1093/mnras/stac3290}

% type= article
\bibitem[{S. {Shahaf} {et~al.}(2025){Shahaf}, {Hallakoun}, {Mazeh}, {Ben-Ami}, {Rekhi}, {El-Badry}, \& {Toonen}}]{Shahaf:2025}
{Shahaf}, S., {Hallakoun}, N., {Mazeh}, T., {et~al.} 2025, \bibinfo{title}{{Correction to: Triage of the Gaia DR3 astrometric orbits. II. A census of white dwarfs},} \mnras, 537, 3594, \dodoi{10.1093/mnras/staf265}

% type= article
\bibitem[{S. {Shahaf} {et~al.}(2019){Shahaf}, {Mazeh}, {Faigler}, \& {Holl}}]{Shahaf:2019}
{Shahaf}, S., {Mazeh}, T., {Faigler}, S., \& {Holl}, B. 2019, \bibinfo{title}{{Triage of astrometric binaries - how to find triple systems and dormant black hole secondaries in the Gaia orbits},} \mnras, 487, 5610, \dodoi{10.1093/mnras/stz1636}

% type= article
\bibitem[{T. {Shenar} {et~al.}(2020){Shenar}, {Bodensteiner}, {Abdul-Masih}, {Fabry}, {Mahy}, {Marchant}, {Banyard}, {Bowman}, {Dsilva}, {Hawcroft}, {Reggiani}, \& {Sana}}]{Shenar:2020}
{Shenar}, T., {Bodensteiner}, J., {Abdul-Masih}, M., {et~al.} 2020, \bibinfo{title}{{The ``hidden'' companion in LB-1 unveiled by spectral disentangling},} \aap, 639, L6, \dodoi{10.1051/0004-6361/202038275}

% type= article
\bibitem[{S.~S. {Shetye} {et~al.}(2024){Shetye}, {Viviani}, {Anderson}, {Mowlavi}, {Eyer}, {Evans}, \& {Szabados}}]{Shetye:2024}
{Shetye}, S.~S., {Viviani}, G., {Anderson}, R.~I., {et~al.} 2024, \bibinfo{title}{{VELOcities of CEpheids (VELOCE): II. Systematic search for spectroscopic binary cepheids},} \aap, 690, A284, \dodoi{10.1051/0004-6361/202450185}

% type= article
\bibitem[{J.~D. {Simon} \& M. {Geha}(2007){Simon} \& {Geha}}]{Simon:2007}
{Simon}, J.~D., \& {Geha}, M. 2007, \bibinfo{title}{{The Kinematics of the Ultra-faint Milky Way Satellites: Solving the Missing Satellite Problem},} \apj, 670, 313, \dodoi{10.1086/521816}

% type= inproceedings
\bibitem[{R.~P. {Stefanik} {et~al.}(1999){Stefanik}, {Latham}, \& {Torres}}]{Stefanik:1999}
{Stefanik}, R.~P., {Latham}, D.~W., \& {Torres}, G. 1999, \bibinfo{title}{{Radial-Velocity Standard Stars},} in Astronomical Society of the Pacific Conference Series, Vol. 185, IAU Colloquium 170: Precise Stellar Radial Velocities, ed. J.~B. {Hearnshaw} \& C.~D. {Scarfe}, 354

% type= article
\bibitem[{M. {Steinmetz} {et~al.}(2020){Steinmetz}, {Guiglion}, {McMillan}, {Matijevi{\v{c}}}, {Enke}, {Kordopatis}, {Zwitter}, {Valentini}, {Chiappini}, {Casagrande}, {Wojno}, {Anguiano}, {Bienaym{\'e}}, {Bijaoui}, {Binney}, {Burton}, {Cass}, {de Laverny}, {Fiegert}, {Freeman}, {Fulbright}, {Gibson}, {Gilmore}, {Grebel}, {Helmi}, {Kunder}, {Munari}, {Navarro}, {Parker}, {Ruchti}, {Recio-Blanco}, {Reid}, {Seabroke}, {Siviero}, {Siebert}, {Stupar}, {Watson}, {Williams}, {Wyse}, {Anders}, {Antoja}, {Birko}, {Bland-Hawthorn}, {Bossini}, {Garc{\'\i}a}, {Carrillo}, {Chaplin}, {Elsworth}, {Famaey}, {Gerhard}, {Jofre}, {Just}, {Mathur}, {Miglio}, {Minchev}, {Monari}, {Mosser}, {Ritter}, {Rodrigues}, {Scholz}, {Sharma}, {Sysoliatina}, \& {RAVE Collaboration}}]{Steinmetz:2020}
{Steinmetz}, M., {Guiglion}, G., {McMillan}, P.~J., {et~al.} 2020, \bibinfo{title}{{The Sixth Data Release of the Radial Velocity Experiment (RAVE). II. Stellar Atmospheric Parameters, Chemical Abundances, and Distances},} \aj, 160, 83, \dodoi{10.3847/1538-3881/ab9ab8}

% type= article
\bibitem[{S.~S. {Vogt} {et~al.}(2014){Vogt}, {Radovan}, {Kibrick}, {Butler}, {Alcott}, {Allen}, {Arriagada}, {Bolte}, {Burt}, {Cabak}, {Chloros}, {Cowley}, {Deich}, {Dupraw}, {Earthman}, {Epps}, {Faber}, {Fischer}, {Gates}, {Hilyard}, {Holden}, {Johnston}, {Keiser}, {Kanto}, {Katsuki}, {Laiterman}, {Lanclos}, {Laughlin}, {Lewis}, {Lockwood}, {Lynam}, {Marcy}, {McLean}, {Miller}, {Misch}, {Peck}, {Pfister}, {Phillips}, {Rivera}, {Sandford}, {Saylor}, {Stover}, {Thompson}, {Walp}, {Ward}, {Wareham}, {Wei}, \& {Wright}}]{Vogt:2014}
{Vogt}, S.~S., {Radovan}, M., {Kibrick}, R., {et~al.} 2014, \bibinfo{title}{{APF{\textemdash}The Lick Observatory Automated Planet Finder},} \pasp, 126, 359, \dodoi{10.1086/676120}

\end{thebibliography}
\bibliographystyle{aasjournalv7}

\end{document}